\documentclass[
  reprint,
  aps,
  pra,
  amsmath,
  amssymb,
  floatfix,
  longbibliography
]{revtex4-2}

\usepackage[utf8]{inputenc} 
\usepackage[T1]{fontenc}
\usepackage[english]{babel}
\usepackage{blindtext}

\usepackage{amsfonts}
\usepackage{amsthm}
\usepackage{mathtools}
\usepackage{stmaryrd}
\usepackage{mathrsfs}
\usepackage{cancel}

\usepackage{physics}

\usepackage{graphicx}
\usepackage{subcaption}
\usepackage{xcolor}
\usepackage{tikz}
\usetikzlibrary{arrows.meta}
\usepackage[section]{placeins}

\usepackage{quantikz}

\usepackage{svg}
\svgpath{{figs/}}

\usepackage{algorithmic}

\newcounter{algorithm}
\renewcommand{\thealgorithm}{\arabic{algorithm}}

\newenvironment{revtexalgorithm}[1][t]
{%
  \begin{figure}[#1]
  \centering
  \refstepcounter{algorithm}%
  \begin{minipage}{0.96\linewidth}
  \hrule height 0.8pt
  \vspace{0.45em}
}
{%
  \vspace{0.45em}
  \hrule height 0.8pt
  \end{minipage}
  \end{figure}
}

\newcommand{\algcaption}[1]{%
  \noindent\textbf{Algorithm~\thealgorithm. #1}\par
  \vspace{0.45em}
  \hrule height 0.4pt
  \vspace{0.45em}
}

\usepackage{soul}
\usepackage{relsize}
\usepackage{gensymb}
\usepackage{wrapfig}
\usepackage{booktabs}

\usepackage[
  breaklinks=true,
  colorlinks=true,
  linkcolor=blue,
  citecolor=blue,
  urlcolor=blue
]{hyperref}

\makeatletter
\def\mathcolor#1#{\@mathcolor{#1}}
\def\@mathcolor#1#2#3{%
  \protect\leavevmode
  \begingroup
    \color#1{#2}#3%
  \endgroup
}
\makeatother

\begin{document}

\title{Trapping Sets of Detector Error Models}
\thanks{ This work is supported by the NSF under grants CCF-2420424, CIF-2106189, CCF-2100013, ECCS/CCSS-2027844, ECCS/CCSS-2052751, and in part by the CoQREATE program under grant ERC-1941583. Bane Vasi\'{c} has disclosed an outside
interest in his startup company QECLabs to The University
of Arizona. Conflicts of interest resulting from this interest are being managed by The University of Arizona in accordance
with its policies.
}%

\author{Michele Pacenti}
\author{Nithin Raveendran}
\author{Bane Vasi\'c}%

\affiliation{
Department of Electrical and Computer Engineering, University of Arizona, Tucson, Arizona, USA.
}

\date{\today}

\begin{abstract}
Message-passing decoders are among the most promising candidates for scalable quantum error correction, yet their behavior in the low-error-rate regime remains poorly understood under realistic circuit-level noise. In this work, we introduce a systematic framework for identifying the graph structures that govern decoder failures and for using them to predict the resulting error floor. We apply exhaustive trapping-set enumeration directly to the detector error model of a bivariate bicycle code and test all low-weight fault configurations supported on the resulting structures. This converts the analysis of extremely rare logical failures into a finite structural search, avoiding the prohibitive cost of direct Monte Carlo simulation.
We evaluate the framework on three iterative decoders with substantially different architectures and decoding heuristics. Remarkably, for \texttt{RelayBP}, the resulting prediction accurately reproduces the simulated error floor; for the others, it remains within the same order of magnitude. Despite their differences, leafless elementary trapping sets capture a substantial part of the low-weight error-floor contribution for all three decoders. Moreover, each decoder admits failures caused by fault configurations well below the correction capability implied by the circuit-level distance, revealing a substantial gap between code distance and practical iterative-decoding performance.
These results establish trapping-set analysis as a practical framework for predicting error floors, exposing the structural weaknesses of iterative decoders, and guiding the joint design of decoding algorithms.
\end{abstract}

\maketitle

\section{Introduction}
Message-passing decoders are among the most promising candidates for quantum low-density parity-check (QLDPC) codes because of their low computational complexity, high degree of parallelism, and architectural simplicity. For several years, belief propagation combined with ordered-statistics decoding (\texttt{BPOSD})~\cite{panteleev_degenerate_2021} represented the state of the art in decoding performance. Its reliance on OSD post-processing, however, introduces a computational cost that scales as $\mathcal O(n^3)$, where $n$ is the code length, making it unsuitable for large-scale fault-tolerant quantum computing. This limitation has motivated the development of alternatives capable of approaching \texttt{BPOSD} performance without expensive post-processing~\cite{yin2024symbreakmitigatingquantumdegeneracy,tsubouchi2026degeneracycuttinglocalefficient}. More recently, several BP-based decoders have been shown to match or even surpass \texttt{BPOSD}. \texttt{RelayBP}~\cite{muller2025improvedbeliefpropagationsufficient} was among the first to demonstrate this behavior and was followed by \texttt{BeamSearch}~\cite{ye2025beamsearchdecoderquantum} and \texttt{Restart\_Belief}~\cite{valentini2025restartbeliefgeneralquantum}, although the latter was not originally benchmarked on circuit-level detector error models. Maan \textit{et al.} subsequently reported similarly strong performance with the \texttt{GARI-NSM} decoder~\cite{Maan2026}, while \texttt{ImpulseBP}~\cite{bhatnagar2026impulsedecodingquantumldpc} has also emerged as a competitive approach with comparatively low complexity.

Despite the many advantages they bring, iterative decoders are still suboptimal, as they generally cannot correct up to the minimum distance of the underlying code. In particular, in the low error regime, it is well-known that they fail for certain small-weight error configurations called \textit{trapping sets} (TS)~\cite{raveendran_trapping_2021}. In the context of classical LDPC codes, TSs have been extensively studied and characterized~\cite{ontology, richardson-stopping-sets, dolecek_as, bani_irregular,bani_regular}, and their knowledge was exploited either to improve the performance of message-passing decoders~\cite{faid1,faid2,6567866} or to design LDPC codes with better error-floor performance~\cite{6169192,Battaglioni2023,6034068}.

In the context of QLDPC codes, the idea of trapping set was first introduced in~\cite{raveendran_trapping_2021}, where the impact of an intrinsically quantum class of harmful objects called \textit{quantum trapping sets} was observed. From there, several works involving TS of QLDPC codes followed up. On the side of decoder design, Pradhan \textit{et al.}~\cite{asit_istc} explicitly used the knowledge of TSs of lifted product codes to design a decoder that combines BP and recurrent neural networks. In a subsequent work~\cite{asit_tinfo}, the same authors carried out a thorough analysis of quantum and classical TS for hypergraph product and lifted product codes, and proposed new BP-based decoders with enhanced performance in the error floor region. In the same spirit, Chytas \textit{et al.}~\cite{chytas_enhanced_2024,chytas2024collectivebitflippingbaseddecoding,chytas2025enhancedminsumdecodingquantum,chytas2026edge} designed several TS-aware iterative decoding algorithms with significant improvement of performance compared to standard BP, and low complexity. We want to stress that all these works have focused on code-capacity error models, while similar work that specifically targets detector error models (DEM) for circuit-level noise is still missing.
A more general approach is developed by Morris \textit{et al.}~\cite{morris2024absorbingsetsquantumldpc}, who analyzed absorbing sets (a particular class of TSs) and derived sufficient conditions under which they induce failures of syndrome-based Gallager-B decoding. Such an analysis is possible because Gallager-B is governed by simple, deterministic local update rules. By contrast, state-of-the-art QLDPC decoders typically incorporate decoder-specific heuristics, including asymmetric message updates, nonuniform processing across the Tanner graph, randomized schedules, and repeated re-initialization. These mechanisms substantially complicate the decoding dynamics and make an analogous analytical characterization prohibitively difficult for practical decoders. This motivates a complementary, decoder-agnostic approach in which harmful structures are identified directly from the observed decoding behavior.

Beverland \textit{et al.}~\cite{beverland2025failfasttechniquesprobe}
study decoding performance in the low-error-rate regime using a different approach. They characterize decoder failures through a failure spectrum, defined as the fraction of weight-$w$ fault configurations that cause a decoding failure. They then combine this information with a search for the smallest harmful fault configurations and with rare-event sampling to estimate logical error rates beyond the range accessible by direct Monte Carlo simulation. Once the model parameters have been obtained, the logical error rate can be evaluated efficiently. While this approach is valid, it does not retain any information on the error patterns that produce decoding failures.

On the other hand, a TS-based approach restricts the analysis
to bounded-size local subgraphs that are plausible generators of the
dominant low-weight failures. Once these structures have been
enumerated, only the fault patterns supported on their small variable-node
sets must be tested. For a fixed maximum TS size, the number
of injected patterns per structure is bounded, and both the structural
search and the decoder evaluations can be parallelized. Furthermore,
the same catalog of trapping sets can be reused to analyze multiple
decoders, and the probabilities of the identified failure mechanisms can
subsequently be evaluated over an arbitrary range of physical error
rates at negligible additional cost. Thus, in the low error regime, the TS-based analysis
provides not only a topological characterization of the decoder's failure mechanisms (which can be used to improve the decoder itself), but also a computationally efficient alternative to Monte Carlo simulation. However, this advantage is necessarily limited
to moderate TS sizes, since the number of candidate structures
itself grows rapidly with the search depth, and it is only possible for message-passing decoders. The advent of AI-based decoders raises the natural question of which types of structures are harmful for them, and definitely constitutes an important line of future work.

These considerations suggest that a thorough characterization of TSs in circuit-level DEMs is needed. Such a characterization would be useful in two ways. First, it could provide an efficient route to predicting the asymptotic behavior of iterative decoders in the low-error-rate regime, without relying exclusively on direct Monte Carlo simulations, in the same spirit as~\cite{beverland2025failfasttechniquesprobe}. Second, it could guide the design of more effective TS-aware decoders. 
However, a general framework for studying TSs of arbitrary DEMs is still missing. Existing approaches, such as those in~\cite{asit_istc,asit_tinfo,chytas_enhanced_2024,chytas2024collectivebitflippingbaseddecoding,chytas2025enhancedminsumdecodingquantum}, are restricted to the code-capacity case, and either rely on prior knowledge of the dominant harmful structures, for example the TSs of the Tanner $(155,64)$ code appearing in Tanner lifted-product constructions, or focus on specific classes of structures rather than providing a full-spectrum characterization. On the other hand, a full characterization of TSs of circuit-level DEMs is particularly challenging: the associated Tanner graphs are large, highly irregular, and may contain a prohibitive number of TSs to enumerate.

\subsection{Our Contribution}

In this work we present a general framework to characterize TSs of circuit-level DEM, and use it to study the failures of several iterative decoders in the literature and get a lower bound for their performance in the low-weight error regime. We draw from the classical LDPC literature, in particular from the work of Hashemi and Banihashemi on exhaustive search algorithms for \textit{leafless elementary} TS (LETS) in LDPC codes~\cite{bani_regular,bani_irregular}, to collect a list of such structures in circuit-level DEM of bivariate bicycle (BB) codes~\cite{bravyi2024high}. While TS enumeration has exponential complexity, finite-length enumeration can be made efficient by exploiting the topological structure of such subgraphs. In particular, TSs can be generated iteratively by starting from short cycles and progressively expanding them with new variable nodes, as observed in~\cite{ontology}. Based on this observation, the analysis in~\cite{bani_irregular} introduces an algorithm, called \textit{dot-path-lollipop search} (\texttt{dpl-search}), which expands short cycles efficiently via a finite set of transformations that is shown to be complete for the class of LETS. For clarity, we describe the algorithm in detail in Appendix~\ref{sec:dpl-search}. The implementation of the \texttt{dpl-search} used in this work, together with the complete dataset of LETSs enumerated for the $[[144,12,12]]$ BB code, is publicly available in~\cite{dem_trapping_sets}. We apply the \texttt{dpl-search} directly on the DEM, obtaining $92,088,583$ LETS of size at most 5. Restricting the search to LETSs substantially reduces the search space by excluding elementary trapping sets with leaves and non-elementary trapping sets. This restriction is well motivated in the classical LDPC setting, where LETSs have been shown to account for the dominant low-weight failure mechanisms of iterative decoders. Whether the same assumption remains valid for QLDPC codes, however, is not clear \textit{a priori}, particularly for circuit-level detector error models, whose Tanner graphs are highly irregular and whose state-of-the-art decoders employ dynamics that differ substantially from those considered in the classical trapping-set literature. Our results show that, despite these differences, restricting the analysis to LETSs is sufficient to obtain accurate error-floor estimates for several classes of circuit-level decoders. This indicates that LETSs capture a substantial part of the dominant low-weight contribution in the error-rate regime considered.
Once we have obtained the list of LETSs, we exhaustively inject errors on their support and enumerate the resulting decoding failures for several decoders. In this way, we can obtain a lower bound for the performance of iterative decoders in the low error regime, without having to run a full Monte Carlo simulation, which would be infeasible for very low error rates. We test this framework for three decoders: \texttt{RelayBP}~\cite{muller2025improvedbeliefpropagationsufficient}, \texttt{ImpulseBP}~\cite{bhatnagar2026impulsedecodingquantumldpc}, and a simple decoder we propose in this paper, which we label as \textit{ensemble layered min-sum}, or \texttt{ELMS}. The choice of these three decoders is deliberate: they represent classes of decoders with significantly different natures. \texttt{RelayBP} is based on sequential, randomized re-initializations, but is inherently local. \texttt{ImpulseBP}, on the other hand, works with an ensemble of decoders, and it is governed by a global processor that modifies each of the decoder's priors, but it is deterministic. Finally, \texttt{ELMS} is also an ensemble decoder, but is randomized, not based on re-initializations and local. By evaluating the error floor estimate of these three different decoders, we can benchmark our analysis on decoders of different types. The results show that for \texttt{RelayBP} the error floor estimate perfectly matches the results obtained with Monte Carlo simulations, while for \texttt{ImpulseBP} and \texttt{ELMS}, the predicted performance slightly underestimates the simulated results, although remaining within the same order of magnitude. By distinguishing the failing LETSs by non-isomorphic classes that we call \textit{topologies}, we show that the top three topologies for number of harmful errors for \texttt{RelayBP} and \texttt{ImpulseBP} are identical, while they remain among the most harmful ones for \texttt{ELMS}. Thus, although the three decoders exhibhit different failing LETSs, they share a common set of particularly vulnerable structures.
This framework serves not only for estimating performance in the error-floor region, but also to capture dominant failure mechanisms of decoders and build an encyclopedia of potentially harmful structures, with the goal of designing better decoders. While we leave this for future work, we highlight that all the decoders analyzed fail for a certain number of weight-four error patterns; considering that the circuit-level distance of the considered $[[144,12,12]]$ code is 12~\cite{bravyi2024high}, and that it can therefore correct (in principle) all errors up to weight 5, it follows that the performance of the analyzed decoders is quite far from the optimal.

The rest of the paper is organized as follows: in Section~\ref{sec:preliminaries} we introduce the necessary preliminaries for linear codes, quantum stabilizer codes and decoders, noise models and message-passing decoders. In Section~\ref{sec:trapping_sets} we formally introduce the notion of trapping set, quantum trapping sets and the \texttt{dpl-search} algorithm. In Section~\ref{sec:ef_estimation} we describe in detail the methodology to obtain the error floor estimate from the list of trapping sets found. In Section~\ref{sec:results} we illustrate the results of our analysis, and in Section~\ref{sec:conclusions} we derive the conclusions of our analysis and motivate future work directions. In Appendix~\ref{sec:dem_structure} we describe in detail the structure of a general circuit-level DEM, while in Appendix~\ref{sec:dpl-search} we describe in detail the \texttt{dpl-search} algorithm.

\section{Preliminaries}
\label{sec:preliminaries}

\subsection{Linear codes and Tanner graphs}
Let $\mathbb{F}_2^{n}$ be the space of length-$n$ vectors with binary coefficients; the \textit{Hamming weight} (or simply weight) of an element in $\mathbb{F}_2^{n}$ is the number of its non-zero entries, and we denote it with $|.|$; we will use the same notation to indicate the cardinality of a set, with the intended meaning clear from context. An $[n,k,d]$ linear code $C \subset \mathbb{F}_2^{n}$ is a linear subspace of $\mathbb{F}_2^{n}$ generated by $k$ elements, such that each nonzero element in $C$ has Hamming weight at least $d$. A code $C$ can be represented by an $(n-k) \times n$ parity check matrix $H$ such that $C = \ker H$. The code rate of $C$ is defined as $R=k/n$. With $C^{\perp} = \mathrm{im}H^T$ we denote dual of the code $C$ with parameters $[n,k^{\perp}=n-k,d^{\perp}]$. If $H$ is \textit{sparse}, \textit{i.e.}, its row and column weights remain constant or grow slowly with the blocklength, the code $C$ is a \textit{low-density parity check} code. A graph $G = (V,E)$ is a collection of vertices $V$ and edges $E$, such that each edge connects two distinct vertices $v_i,v_j$ and can be represented by the pair $(v_i,v_j)$. A bipartite graph $G = (U \cup V, E)$ is a graph where the vertices can be partitioned in two sets, and the edges only go from one partition to the other. Let $U,V$ be the sets of left and right vertices, respectively. To a bipartite graph is associated a biadjacency matrix $\mathbf{M} \in \mathbb{F}_2^{|U|\times |V|}$; each entry of the matrix $m_{i,j}=1$ if there exists an edge between the $i$-th vertex in $U$ and the $j$-th vertex in $V$, and $m_{i,j}=0$ otherwise.  The parity check matrix is the biadjacency matrix of the \textit{Tanner~graph} $G=(V\cup C,E)$, where the nodes in $V$ are called \textit{variable nodes} and constitute the set of left vertices, the nodes in $C$ are called \textit{check nodes} and constitute the set of right vertices, and there is an edge between $v_j \in V$ and $c_i \in C$ if $h_{ij}=1$, where $h_{ij}$ is the element in the $i$-th row and $j$-th column of $H$. The \textit{degree} of a node is the number of incident edges to that node. If all the variable (check) nodes have the same degree we say the code has \textit{regular} variable (check) degree, and we denote it with $d_v$ ($d_c)$. A \textit{cycle} is a closed path in the Tanner graph, and we denote its length by the number of variable and check nodes in the cycle. The \textit{girth} $g$ of a Tanner graph is the length of its shortest cycle. 

\subsection{Quantum stabilizer codes and decoding}
Let
\begin{equation*}
    \mathcal P_n \coloneqq \{\pm1, \pm i\} \cdot \{I,X,Y,Z\}^{\otimes n}
\end{equation*}
be the $n$-fold Pauli group and $(\mathbb{C}^2)^{\otimes n}$ be the $2^n$-dimensional Hilbert space. An $\llbracket n,k,d \rrbracket$ stabilizer code $\mathcal C$ is defined by an Abelian subgroup $\mathcal S \subset \mathcal P_n$ such that $-I \not \in \mathcal S$. The code space is the common $+1$ eigenspace of all Pauli operators in $\mathcal S$:
\begin{equation*}
    \mathcal C = \{\ket{\psi} \in (\mathbb{C}^2)^{\otimes n} : S\ket{\psi}=\ket{\psi}, \forall S\in \mathcal S\}.
\end{equation*}
Let $S_1,...,S_{n-k}$ be the stabilizer group generators such that $\mathcal S = \langle S_1,...,S_{n-k} \rangle$; since they are all linearly independent and commuting, it holds that $\dim (\mathcal C) = 2^{k}$. The centralizer of $\mathcal S$ in $\mathcal P_n$, denoted as $N(\mathcal S)$, contains all Pauli operators that commute with elements of $\mathcal S$. The logical operators of the code are defined as
\begin{equation*}
    \mathcal L \cong N(\mathcal S) / \mathcal S,
\end{equation*}
namely, the subgroup of Pauli matrices that commute with the stabilizers that are not stabilizers themselves. Logical operators generally act non-trivially on the encoded state, except for the identity operator. The weight of an element of the Pauli group, denoted as $|.|$ is defined as the number of non-identity elements in its tensor product, and the minimum distance of the code is defined as the minimum weight of a logical operator:
\begin{equation*}
    d \coloneqq \min_{L \in \mathcal L} |L|.
\end{equation*}

Let $E \in \mathcal P_n$ be a Pauli error acting on a codeword $\ket{\psi}$. One can define a syndrome map $\sigma : \mathcal P_n \to \mathbb{Z}_2^{n-k}$ as:
\begin{equation*}
    \sigma(E)_i = \left\{\begin{array}{c c}
       0  & \mathrm{if}\ [E,S_i]=0 \\
        1 & \mathrm{if}\ \{E,S_i\}=0,
    \end{array} \right.
\end{equation*}
where $[.,.]$ and $\{.,.\}$ denote the commutator and anti-commutator operators. In other words, the $i$-th bit of the syndrome is $0$ if the error commutes with the $i$-th stabilizer, and is $1$ if the error anti-commutes with the $i$-th stabilizer. One can show that $\sigma(E)$ effectively corresponds to an error syndrome which is then used for decoding.
For each syndrome $ s$ we define the \textit{error coset} $\mathcal E_{ s}$ as the set of all the Pauli operators producing that syndrome:
\begin{equation*}
    \mathcal E_{ s} \coloneqq \{E \in \mathcal P_n: \sigma(E) =  s\}.
\end{equation*}
Let $E_0 \in \mathcal E_{ s}$ be a particular error in the coset, then the entire coset can be written as
\begin{equation*}
    \mathcal E_{ s} =  E_0N(\mathcal S) = \bigsqcup_{L \in \mathcal L, S\in \mathcal S} E_0LS.
\end{equation*}
Thus, we can further partition $\mathcal E_{ s}$ into $2^{2k}$ \textit{logical cosets} by fixing $L$, thus obtaining:
\begin{equation*}
    \mathcal E_{ s}^{(L)} \coloneqq E_0L\mathcal S = \bigsqcup_{ S\in \mathcal S} E_0LS.
\end{equation*}
Crucially, each error in $\mathcal E_{ s}^L$ has the same effect on the encoded state, therefore given an error $E \in \mathcal E_{ s}^{(L)}$, any recovery operator $\hat{E}\in \mathcal E_{ s}^{(L)}$ constitutes a valid error estimate, as the final state after recovery is:
\begin{equation*}
    \hat{E}^{\dagger}E\ket{\psi} = (E_0LS')^{\dagger}(E_0LS)\ket{\psi} = S'^{\dagger}S\ket{\psi} = \ket{\psi}.
\end{equation*}
Clearly, if $E$ and $\hat{E}$ belong to different logical cosets, the resulting state is $L'^{\dagger}L\ket{\psi} \neq \ket{\psi}$, thus resulting in a \textit{logical error}.

A decoder is a map:
\begin{equation*}
    \Lambda : \mathbb F_2^{n-k} \to \mathcal P_n
\end{equation*}
which, given a syndrome $ s$, outputs an estimated recovery operator $\hat{E}$. As just mentioned, the decoder is \textit{successful} if $E$ and $\hat{E}$ belong to the same logical coset, namely, $\hat{E}^{\dagger}E \in \mathcal S$. Given a probabilistic error model $Pr(E)$ on the Pauli group, decoding is therefore a statistical inference problem over logical cosets. The optimal strategy to minimize the probability of error is maximum-likelihood coset decoding. This involves selecting the logical class $L$ that maximizes the total probability of its corresponding logical coset:
\begin{equation*}
\hat{L}_{\rm opt} = \underset{L \in \mathcal L}{\arg \max}\ Pr(\mathcal E_{\mathbf{s}}^{(L)} |  s),
\label{eq:mldc}
\end{equation*}
where
\begin{equation*}
   Pr(\mathcal E_{\mathbf{s}}^{(L)} |  s) = \sum_{S_i \in \mathcal S} Pr(LS_iE_0 |  s),
\end{equation*}
with $E_0$ being an error matching the syndrome.
Notice that once the logical $L$ is identified, the recovery operator is simply any Pauli operator in $\mathcal E_{\mathbf{s}}^{(L)}$. 
In practice, exact maximum-likelihood coset decoding is generally
intractable, since it requires summing over all stabilizer elements in each
logical coset. A common approximation is therefore to replace
maximum-likelihood coset decoding by a minimum-weight criterion. Let
$\sigma(E)$ denote the syndrome of a Pauli error $E$, and let $w(E)$ denote
its Pauli weight. The minimum-weight decoder selects an error of smallest
weight among all errors compatible with the measured syndrome:
\begin{equation*}
    \hat E_{\rm MW}
    \in
    \underset{E' \in \mathcal P_n:\ \sigma(E')= s}{\arg\min}
    \ w(E') .
    \label{eq:mwd}
\end{equation*}
Equivalently, fixing any representative $E_0$ with syndrome $ s$, every
compatible error can be written as $LS_iE_0$, with $L \in \mathcal L$ and
$S_i \in \mathcal S$. Hence, one may define the syndrome-dependent weight of
a logical class as
\begin{equation*}
    d_{ s}(L)
    =
    \min_{S_i \in \mathcal S} w(LS_iE_0),
\end{equation*}
and choose
\begin{equation*}
    \hat L_{\rm MW}
    \in
    \underset{L \in \mathcal L}{\arg\min}
    \ d_{ s}(L).
    \label{eq:mwcoset}
\end{equation*}
The corresponding recovery can then be chosen as any Pauli operator in
$\mathcal E_{ s}^{(\hat L_{\rm MW})}$ attaining this minimum.

For independent Pauli noise with sufficiently small physical error rate,
lower-weight errors are exponentially more likely than higher-weight errors.
In this regime, minimum-weight decoding provides a natural approximation to
maximum-likelihood decoding.

It is convenient to represent Pauli operators using the binary symplectic representation. Consider the map $\phi : \mathcal P_n \longrightarrow \mathbb F_2^{2n}$, defined modulo the phase subgroup $\{\pm 1, \pm i\}$.
For $P = P_1 \otimes \cdots \otimes P_n \in \mathcal P_n$, its image is the vector
$$
\phi(P) = ( x \mid  z) \in \mathbb F_2^{2n},
$$
where $ x,  z \in \mathbb F_2^n$ are defined componentwise by
$$
x_i =
\begin{cases}
1 & \text{if } P_i \in \{X,Y\},\\
0 & \text{otherwise},
\end{cases}
\qquad
z_i =
\begin{cases}
1 & \text{if } P_i \in \{Z,Y\},\\
0 & \text{otherwise}.
\end{cases}
$$
Under this correspondence, multiplication of Pauli operators modulo phase corresponds to vector addition in $\mathbb F_2^{2n}$, and the commutation relation is determined by the symplectic inner product
$$
 u \odot {v} \coloneqq  v_x \cdot  u_z +  v_z \cdot  u_x
\mod 2.
$$
It follows that the stabilizer generators $S_1,...,S_{n-k}$ can be represented with a parity check matrix ${H} = [{H}_X\ {H}_Z] \in \mathbb{F}_2^{(n-k)\times 2n}$, and the syndrome ${s} = \sigma(E)$ can be obtained as:
\begin{equation*}
     s = {e} \odot  {H}^T,
\end{equation*}
where ${e}$ is the symplectic representation of the Pauli error $E$.

An $\llbracket n, k_X-k_Z^{\perp}, d \rrbracket $ Calderbank-Shor-Steane (CSS) code $\mathcal{C}$ is a stabilizer code constructed using two classical  $[n,k_X,d_X]$ and $[n,k_Z,d_Z]$ codes $C_X = \ker H_X$ and $C_Z = \ker H_Z$, respectively, such that $C_Z^{\perp} \subset C_X$ and $C_X^{\perp} \subset C_Z$ \cite{calderbank_good_1996}. The minimum distance is $d=~\mathrm{min}\{d_X,d_Z\}$, with $d_X$ being the minimum Hamming weight of a codeword in $C_X \setminus C_Z^{\perp}$, and $d_Z$ being the minimum Hamming weight of a codeword in $C_Z \setminus C_X^{\perp}$. Notice that if $H_XH_Z^T=\mathbf{0}$ it immediately follows that $C_X^{\perp} \subset C_Z$ and that $C_Z^{\perp} \subset C_X$, thus it is sufficient to satisfy such constraint when designing the two parity check matrices $H_X$ and $H_Z$ to obtain a valid CSS code.
The parity check matrix of a CSS code has the form:
\begin{equation*}
    H = \left[\begin{array}{c c}
         H_X &  0\\
          0 &  H_Z
    \end{array} \right],
\end{equation*}
which allows, in principle, to decode $X$ and $Z$ errors separately. Thus, w.l.o.g., throughout the rest of this paper we assume to only decode $X$ errors, such that $H =  H_Z$, and the decoding problem can be reformulated in the following way. Let $m$ be the number of rows of $ H$; we define our decoder to be the map:
\begin{equation*}
    \Lambda : \mathbb{F}_2^{m} \to \mathbb F_2^{n}
\end{equation*}
which, given a syndrome $ s =  e H^T$, outputs an estimated $\hat{ e}$ such that $ e + \hat{ e} \in \mathrm{rowspace}( H_X)$, and (\ref{eq:mldc}) remains the same, except that $\mathcal L$ becomes $\mathcal L_X$, which is the set of $X$ logical operators.

\subsection{Noise models, syndrome extraction and detectors}

In the standard code capacity model, each data qubit is assumed to undergo depolarizing noise, whose action can be described by
\begin{equation*}
    \mathcal D(\rho) = (1-p)I\rho I + \frac{p}{3}(X\rho X + Y\rho Y + Z\rho Z),
\end{equation*}
where $\rho$ is the density operator corresponding to the quantum state. In other words, each data qubit has a probability of an $X$, $Y$ or $Z$ error with probability $p/3$ each. The syndrome is computed through a syndrome extraction circuit. This circuits consists of preparing ancilla qubits in the $\ket{0}$ state for $Z$-type stabilizers and $\ket{+}$ state for $X$-type stabilizers, and performing a sequence of CNOT gates between data and ancilla qubits to extract the syndrome.
However, in a realistic setting, every component in this circuit is noisy. In particular:
\begin{itemize}
    \item Each ancilla qubit is intended to be initialized in the $\ket{0}$ (or $\ket{+}$) state. With probability $p$, the orthogonal state $\ket{1}$ (or $\ket{-}$) is prepared instead.
    \item The measurement of an ancilla qubit is flipped with probability $p$. This corresponds to a classical bit-flip error on the measurement outcome $m \to m \oplus 1$.
    \item Each single-qubit unitary gate is followed by a single-qubit depolarizing channel with error probability $p$. Similarly, every two-qubit gate (such as CNOT or CZ) is followed by a two-qubit depolarizing channel. In the latter case, with probability $p$, one of the 15 non-identity two-qubit Pauli operators $P \in \{I, X, Y, Z\}^{\otimes 2} \setminus \{I \otimes I\}$ is applied.
    \item  Any qubit (data or ancilla) that is not involved in a gate during a time step undergoes a single-qubit depolarizing channel with probability $p$.
\end{itemize}

Crucially, circuit-level noise cannot be directly decoded using the original Tanner graph of the code for two main reasons: $i)$ the decoder would fail to account for space-time correlated error mechanisms (hook errors), and $ii)$ the noise introduces syndrome outcomes that lie outside the column space of the original parity check matrix $H$. Consequently, the code under circuit-level noise can be effectively modeled as a new code, referred to as the \textit{outcome code} or \textit{detector error model} (DEM) in literature~\cite{fault_complex}. We give a detailed description of a general DEM in Appendix~\ref{sec:dem_structure}.

\subsection{Message-passing decoders}
Message-passing decoders are one of the major classes of decoders for QLDPC codes.  In this subsection, we introduce the Min-Sum (MS) decoder, which constitutes a low-complexity approximation of the BP decoder. The MS decoder is an iterative algorithm.
In each iteration, messages between variable nodes and check nodes are exchanged. A check-to-variable message sent from check node $i$ to variable node $j$ at the $\ell$-th iteration is denoted by $\mu_{i,j}^{(\ell)}$, whereas a variable-to-check message sent from variable node $j$ to check node $i$ at the $\ell$-th iteration is denoted by $\nu_{j,i}^{(\ell)}$. The variable node update rule is computed as:
\begin{equation*}
    \nu_{j,i}^{(\ell)} = \lambda_j + \sum_{i' \in \mathcal N(j)\setminus i} \mu_{i',j}^{(\ell)},
    \label{eq:v2c}
\end{equation*}
where $\lambda_j$ is the prior \textit{log-likelihood ratio} (LLR) for variable node $j$ and $\mathcal N(j)$ is the set of neighbors of the node $j$. The check node update rule is computed as:
\begin{equation*}
   \mu_{i,j}^{(\ell)} = (1-2s_i)\prod_{j'\in \mathcal N(i)\setminus j} \mathrm{sgn}( \nu_{j',i}^{(\ell-1)}) \cdot \underset{j'\in \mathcal N(i)\setminus j}{\min} \abs{\nu_{j',i}^{(\ell-1)}},
   \label{eq:c2v}
\end{equation*}
where $s_i$ is the $i$-th syndrome bit. Here, the sign function is defined as
\begin{equation*}
    \mathrm{sgn}(u) = \begin{cases}
        -1,\  \mathrm{if}\ u<0\\
        +1,\ \rm otherwise.
    \end{cases}
\end{equation*}

At the end of each iteration, the error estimate $\hat{\mathbf{e}}$ is given by the decision update function:
\begin{equation*}
    \hat{e}_j = \frac
    {1-\mathrm{sgn}\left(\lambda_j + \sum_{i \in \mathcal N(j)} \mu_{i,j}^{(\ell)}\right)}{2}.
    \label{eq:decision}
\end{equation*}
The decoding procedure is continued until the maximum number of iterations has been reached or until the syndrome is matched by $\hat{\mathbf{e}}$.

\subsection{Ensemble Layered Min-Sum}
In this work, we also introduce a decoder intended primarily as a benchmark for our framework, which we call \textit{ensemble layered min-sum} (\texttt{ELMS}). Although \texttt{ELMS} generally underperforms \texttt{RelayBP} and \texttt{ImpulseBP}, it is considerably simpler and faster, while still achieving competitive performance relative to methods such as \texttt{SymBreak}~\cite{yin2024symbreakmitigatingquantumdegeneracy} and \texttt{degeneracy cutting}~\cite{tsubouchi2026degeneracycuttinglocalefficient}.

The \texttt{ELMS} decoder consists of an ensemble of 20 layered min-sum decoders. In each constituent decoder, the check nodes of the detector error model (DEM) are partitioned into layers such that no two check nodes within the same layer share a common neighboring variable node. During decoding, all check nodes in a given layer are updated in parallel, followed by the update of all variable nodes adjacent to at least one check node in that layer. The procedure is then repeated for the subsequent layers. One decoding iteration is completed once every layer has been processed. Consequently, a variable node connected to check nodes belonging to different layers may be updated multiple times within the same iteration.

To introduce diversity across the ensemble and reduce the dependence on a particular layered schedule, the order in which the layers are processed is randomized independently at each iteration. Each constituent decoder performs at most 100 iterations. Once all 20 decoders have terminated, the final error estimate is selected as the minimum-weight estimate among those produced by the decoders that have successfully converged.
We additionally apply damping to the variable-to-check messages. In particular, the message sent from variable node \(j\) to check node \(i\) at step \(\ell\) is given by
\begin{equation*}
    \nu_{j,i}^{(\ell)}
    =
    \alpha
    \left(
        \lambda_j
        +
        \sum_{i' \in \mathcal N(j)\setminus\{i\}}
        \mu_{i',j}^{(\ell)}
    \right)
    +
    (1-\alpha)\nu_{j,i}^{(\ell-1)},
    \label{eq:v2c}
\end{equation*}
where the damping parameter is set to \(\alpha=0.95\). Notice that since the same variable node is updated multiple times during the same iteration, here $\ell-1$ refer to the previous message sent, which may occur within the same iteration.

\section{Trapping Sets}
\label{sec:trapping_sets}
In this Section, we formally introduce the notion of TS. First, we define classical TSs, giving a brief historical context, and later we define quantum TSs (QTSs). Finally, we describe the problem of TS enumeration.

\subsection{Classical Trapping Sets}

Let $G=(V\cup C,E)$ be a Tanner graph. For a subset $S\subseteq V$, let
$G(S)$ denote the subgraph induced by the variable nodes in $S$ and all check
nodes adjacent to at least one variable node in $S$. The set of check nodes
adjacent to $S$ is denoted by $\mathcal N(S)$.
The check nodes in $\mathcal N(S)$ are partitioned according to the parity of
their degree in the induced subgraph $G(S)$. In particular, we define

$$
\mathcal{N}_o(S)
=
\left\{
c\in\mathcal{N}(S):
\deg_{G(S)}(c)\text{ is odd}
\right\},
$$

and

$$
\mathcal{N}_e(S)
=
\left\{
c\in\mathcal{N}(S):
\deg_{G(S)}(c)\text{ is even}
\right\}.
$$
A
trapping set belonging to the class $(a,b)$ is a subgraph $G(S)$ such that:
$$
|S|=a,
\qquad
|\mathcal{N}_o(S)|=b.
$$

We emphasize that the term ``trapping set'' is deliberately generic: at its most basic level, it denotes a set of variable nodes whose associated error pattern causes, or is strongly correlated with, a decoding failure. Because such failures are often governed by the local graph structure induced by the error support, trapping sets have historically been studied through combinatorial characterizations. Depending on the channel model, the code ensemble, and the decoder, related harmful configurations appear under different names. For the binary erasure channel, the relevant objects are \emph{stopping sets}~\cite{richardson-stopping-sets}, namely sets of variable nodes whose neighboring check nodes all have induced degree at least two. Such configurations are precisely the obstruction to iterative peeling decoding, since no degree-one check is available from which to recover an erased variable.

In other decoding contexts, particularly for bit-flipping and related iterative decoders, a prominent role is played by \emph{absorbing sets}~\cite{dolecek_as}, sometimes also referred to as fixed sets. An absorbing set is a trapping-set-like configuration in which every variable node in the set is connected to more satisfied checks than unsatisfied checks. This stability condition captures the fact that the decoder receives stronger local evidence in favor of keeping the erroneous variables unchanged than of flipping them. A stronger notion is that of a \emph{fully absorbing set}, where one additionally requires the variable nodes outside the set to be locally stable, in the sense that they also see more satisfied checks than unsatisfied checks.

In this paper, we focus instead on \emph{leafless elementary trapping sets} (LETSs). For an elementary trapping set (ETS), every check node in $G(S)$ has induced degree
one or two. Therefore, $\mathcal{N}_o(S)$ consists precisely of the degree-one
check nodes, whereas $\mathcal{N}_e(S)$ consists of the degree-two check nodes.
A key advantage of ETSs is that they admit a compact line-and-point, or normal-graph, representation: variable nodes are represented as points, while degree-two check nodes are replaced by edges between the corresponding variables. Degree-one checks are then omitted. This representation exposes the internal graph structure of the trapping set and allows ETSs to be organized hierarchically into an \emph{ontology}~\cite{ontology}, in which larger trapping sets are obtained by expanding or combining smaller ones. This hierarchical structure is one of the main reasons ETSs are well suited for systematic enumeration and for decoder- and code-design analyses. A LETS is therefore an ETS whose normal graph does not possess any leaf (in other words, each variable node is connected to at least two others).

\subsection{Quantum Trapping Sets}
In the context of QLDPC codes, a distinguished class of trapping sets, called
\emph{quantum trapping sets} (QTSs)~\cite{raveendran_trapping_2021}, has been
shown to be detrimental to iterative decoders. A QTS is the subgraph induced by
the support of a stabilizer of the code. Since stabilizers have trivial
syndrome, a QTS is naturally classified as an \((a,0)\) trapping set, where
\(a\) denotes the weight of the corresponding stabilizer. Notice that, in principle, an $(a,0)$ QTS can also be a logical operator, if $a$ is equal to the minimum distance of the code; however, the definition of trapping set implies implicitly that $a$ is small compared to the minimum distance of the code, therefore we can assume that QTS are induced by supports of stabilizers without loss of generality.

The relevance of QTSs is tied to quantum degeneracy. Indeed, let \(S\) be the
support of a stabilizer, with \(|S|=a\). For any error pattern \(e_1\) supported
on \(S\), there exists a complementary error pattern
\[
e_2 = e_1 + \mathbf{1}_S
\]
such that \(e_1\) and \(e_2\) have the same syndrome, since they differ by a
stabilizer. If \(\operatorname{wt}(e_1)=w\), then
\(\operatorname{wt}(e_2)=a-w\), assuming both errors are supported on \(S\).
Thus, a QTS describes a pair, or more generally a family, of degenerate error
configurations that are indistinguishable from the syndrome alone.

This phenomenon has no direct analogue in the classical case. A classical
\((a,0)\) trapping set is simply the support of a codeword, whereas in the
quantum setting a zero-syndrome stabilizer support represents a degeneracy of
the error model: two errors differing by that stabilizer correspond to the same
error coset. In particular, when \(a\) is even and
\(\operatorname{wt}(e_1)=\operatorname{wt}(e_2)=a/2\), the two degenerate
configurations are equally likely under a symmetric independent noise model.
Message-passing decoders may then fail to converge, or converge unreliably,
because the local information passed on the Tanner graph does not distinguish
between these competing degenerate solutions.



While QTS have been characterized in the context of code-capacity noise models, we highlight that even for circuit-level DEM the same definition holds. Let \(D\) denote the detector matrix of the circuit-level model, and let
\(L\) denote the matrix mapping faults to logical observables. A set of circuit
faults \(F\) is a stabilizer of the DEM if it produces neither detector events nor a
logical fault, namely
\[
D\mathbf{1}_F^T = 0,
\qquad
L\mathbf{1}_F^T = 0.
\]
Equivalently, \(F\) belongs to the kernel of the combined map
\[
\begin{bmatrix}
D \\ L
\end{bmatrix}.
\]
Such a fault set is invisible both to the detector syndrome and to the logical
observables, and therefore plays in the DEM the same degeneracy-inducing role
that stabilizer supports play in the code-capacity Tanner graph. This motivates
the study of DEM trapping structures as zero-detector, zero-logical fault
configurations, and suggests that QTS analysis can be extended beyond
code-capacity and phenomenological noise to the full circuit-level setting.

\subsection{Trapping Set Search}
Since both classical and quantum trapping sets may constitute valid failure configurations for message-passing decoders, it is important to characterize both. For the purpose of enumeration, however, it is convenient to describe quantum trapping sets simply as $(a,0)$ trapping sets. They can then be searched for using essentially the same combinatorial framework developed for classical trapping sets. Indeed, the classical LDPC literature contains a variety of efficient techniques for enumerating trapping sets~\cite{6192273,bani_regular}.

Several approaches can be used for this purpose. One relatively direct method, and one that is naturally applicable to QLDPC codes, is the procedure proposed by Raveendran \textit{et al.}~\cite{raveendran_expansion}. The algorithm first constructs clusters by expanding short cycles with adjacent variable nodes up to a predetermined size; this is the expansion step. Then, errors are exhaustively injected within the expanded subgraph, and the decoder is run on each resulting error pattern. Whenever decoding fails, the union of the supports of the failing configurations is identified as a trapping set; this is the contraction step. Although this procedure is heuristic and not exhaustive, it has the important advantage of making very few structural assumptions. In particular, it does not rely on a specific combinatorial characterization of the trapping set, nor on restrictive properties of the Tanner graph such as regular variable degree or fixed girth. This makes the method broadly applicable.
However, when attempting to apply this technique to DEMs arising from circuit-level noise, we encountered substantial difficulties. The Tanner graphs associated with these DEMs are typically highly connected and contain check nodes of relatively large degree. As a result, the expansion step rapidly produces clusters whose size is too large for the subsequent exhaustive contraction step. In practice, this renders the algorithm infeasible for circuit-level DEMs, unless the expansion rule is redesigned to exploit the specific graphical structure of the DEM.

Another class of trapping-set enumeration methods relies on combinatorial search. In this work, we consider the \texttt{dpl-search} algorithm introduced in~\cite{bani_irregular} and apply it to circuit-level DEMs. The complete algorithm is described in Appendix~\ref{sec:dpl-search}, while here we illustrate it qualitatively, illustrating its pseudocode in Algorithm~\ref{alg:qualitative}.
The DPL algorithm builds LETS starting from cycles in the Tanner graph. It consists of four main subroutines: \texttt{enumerate\_cycles}, \texttt{dot\_expansion}, \texttt{path\_expansion}, and \texttt{lollipop\_expansion}. The search first enumerates cycles whose sizes are compatible with the chosen maximum trapping-set size $a_{\max}$. Each cycle is then treated as an initial trapping set and is progressively enlarged by applying one of three elementary expansion operations. At every stage, only structures satisfying the bound on the number of unsatisfied checks are retained.

In a \texttt{dot\_expansion}, a single variable node is added to the current trapping set. The new variable node must be connected to at least two check nodes already contained in the induced subgraph, so that the resulting structure remains a valid elementary trapping set.

In a \texttt{path\_expansion}, several new variable nodes are added at once. These nodes form a path whose two endpoints connect to check nodes of the current trapping set. The internal check nodes of the path are new, while the endpoints attach the path to the existing structure.

The \texttt{lollipop\_expansion} combines a path with an additional cycle. Starting from a check node of the current trapping set, a path is added that terminates on a new cycle. This operation therefore appends a cycle to the existing structure through a connecting path, producing a larger and more complex trapping set.

\begin{revtexalgorithm}
\algcaption{Qualitative description of dpl-search}
\label{alg:qualitative}
\begin{algorithmic}[1]
\REQUIRE Tanner graph $G$, maximum trapping-set size $a_{\max}$,
bounds $b_{\max}^{a}$, and expansion table $\mathrm{EX}$
\ENSURE All LETS instances within the prescribed $(a,b)$ range

\STATE $\mathcal{I}\gets\emptyset$

\COMMENT{Step 1: Find the cycles that will be used as starting structures}

\FOR{each cycle size $k=g/2,\ldots,a_{\max}$}
    \STATE $\mathcal{C}_{k}\gets \texttt{enumerate\_cycles}(G,k)$
    \STATE Add all cycles in $\mathcal{C}_{k}$ to $\mathcal{I}$
\ENDFOR

\COMMENT{Step 2: Grow each cycle into larger trapping sets}

\FOR{each initial cycle size $k=g/2,\ldots,a_{\max}$}

    \STATE Let $\mathcal{S}$ be the set of structures generated from
    cycles of size $k$

    \FOR{$a=k,\ldots,a_{\max}-1$}

        \FOR{each trapping set $T\in\mathcal{S}$ with parameters $(a,b)$}

            \STATE Read from $\mathrm{EX}(a,b)$ which expansions
            should be applied to $T$

            \IF{a dot expansion is listed in $\mathrm{EX}(a,b)$}
                \STATE $\mathcal{N}\gets
                \texttt{dot\_expansion}(G,T)$
                \STATE Add the valid structures in $\mathcal{N}$
                to $\mathcal{S}$ and $\mathcal{I}$
            \ENDIF

            \FOR{each allowed path length $m$ listed in
            $\mathrm{EX}(a,b)$}
                \STATE $\mathcal{N}\gets
                \texttt{path\_expansion}(G,T,m)$
                \STATE Add the valid structures in $\mathcal{N}$
                to $\mathcal{S}$ and $\mathcal{I}$
            \ENDFOR

            \FOR{each allowed lollipop expansion
            $(c,m)$ listed in $\mathrm{EX}(a,b)$}
                \STATE $\mathcal{N}\gets
                \texttt{lollipop\_expansion}(G,T,\mathcal{C}_{c},m)$
                \STATE Add the valid structures in $\mathcal{N}$
                to $\mathcal{S}$ and $\mathcal{I}$
            \ENDFOR

        \ENDFOR

        \STATE Remove duplicate structures from $\mathcal{S}$

        \STATE Discard every structure with size larger than
        $a_{\max}$ or with $b>b_{\max}^{a}$

    \ENDFOR
\ENDFOR

\STATE \textbf{return} $\mathcal{I}$
\end{algorithmic}
\end{revtexalgorithm}

These three operations form a complete set of primitive expansions for generating all LETSs within the fixed values of $a$ and $b$. However, blindly applying every possible expansion to every intermediate structure would be computationally inefficient. The DPL algorithm therefore uses an expansion table, denoted by $\mathrm{EX}$, which specifies which dot, path, or lollipop expansions are required for a trapping set with given parameters $(a,b)$. This table is generated using Algorithm 1 in~\cite{bani_irregular}, given $b_{\rm max}$ and the variable degrees list in the Tanner graph. Notice that intermediate structures generated by the algorithm may have $b>b_{\rm max}$; let $b^a_{\rm max}$ be the maximum value of $b$ of an intermediate structure $a<a_{\rm max}$; the value of $b^a_{\rm max}$ is calculated given the degree distribution of the variable nodes in the Tanner graph, and the explicit formula can be found in~\cite{bani_irregular}. Starting from each initial cycle, the algorithm follows this table and repeatedly applies the allowed expansions until the maximum size $a_{\max}$ is reached. Newly generated trapping sets are stored according to their values of $a$ and $b$, duplicates are removed, and only structures satisfying the corresponding bound $b\leq b_{\max}^{a}$ are retained. In this way, the algorithm avoids unnecessary transformations while still exhaustively enumerating all trapping sets in the target range.

We apply Algorithm~\ref{alg:qualitative} to the circuit-level DEM. In particular we consider bivariate bicycle codes with their corresponding syndrome extraction circuits proposed in~\cite{bravyi2024high}. We report in Table~\ref{tab:ets-counts} the enumeration of the LETS structures found with this criterion. We set $a_{\rm max}=5$ and $b_{\rm max}=5$, as searching for larger parameters leads to an overwhelming number of structures that overflows the memory. We show empirically that our search still captures the dominant failures for several decoders, and that the found structures can be utilized to obtain a relatively tight lower bound on the error floor performance.

\begin{table}[t]
    \centering
    \begin{tabular}{crrrrr}
        \toprule
        $b\backslash a$ & $2$ & $3$ & $4$ & $5$ & Total \\
        \midrule
        $0$ & -- & 8\,640 & 39\,995 & 194\,457 & 243\,092 \\
        $2$ & 5\,040 & 20\,918 & 176\,256 & 1\,257\,912 & 1\,460\,126 \\
        $3$ & 6\,048 & 53\,826 & 524\,088 & 6\,824\,502 & 7\,408\,464 \\
        $4$ & 6\,693 & 62\,649 & 1\,398\,780 & 22\,244\,970 & 23\,713\,092 \\
        $5$ & 9\,360 & 88\,056 & 2\,496\,600 & 56\,669\,793 & 59\,263\,809 \\
        \midrule
        Total & 27\,141 & 234\,089 & 4\,635\,719 & 87\,191\,634 & 92\,088\,583 \\
        \bottomrule
    \end{tabular}
    \caption{Number of LETSs found for each $(a,b)$ class.}
    \label{tab:ets-counts}
\end{table}

\section{Error floor estimation}
\label{sec:ef_estimation}

Once the list of trapping sets has been obtained, we exhaustively inject
fault patterns of weights one through four on their variable-node
supports and record the resulting decoding outcomes. No weight-one or
weight-two failures were observed for any of the three decoders; \texttt{RelayBP} and \texttt{ImpulseBP} exhibit failures at weight four, while \texttt{ELMS} fails also for weight three errors. Contributions from
weight-five patterns are smaller and are
asymptotically negligible in the low-error-rate regime considered here.

Let \(\mathcal{T}\) denote the set of enumerated LETS instances. We
define the set of distinct weight-\(w\) fault supports contained in at
least one enumerated LETS as
\begin{equation*}
\mathcal{U}_w
=
\bigcup_{T\in\mathcal{T}}
\left\{
S\subseteq \operatorname{supp}(T):
|S|=w
\right\},
\label{eq:test-support-universe}
\end{equation*}
with $w\in\{1,2,3,4\}$.
Taking the union of the supports avoids double counting fault patterns
that occur in more than one LETS, for instance when one LETS is
contained in another.
Let \(p\) be the physical error rate, and let \(i=1,\ldots,N\) index
the independent fault mechanisms in the detector error model. We
denote by \(\pi_i(p)\) the probability of fault mechanism \(i\), as
obtained from the circuit at physical error rate \(p\). The probability
that exactly the faults in \(S\) occur is
\begin{equation}
\Pr(S;p)
=
\prod_{i\in S}\pi_i(p)
\prod_{j\notin S}
\left[1-\pi_j(p)\right].
\label{eq:fault-support-probability}
\end{equation}
Distinct fault supports describe mutually exclusive physical events.

Let \(N_a\) be the number of enumerated LETS instances of size \(a\),
as reported in Table~\ref{tab:ets-counts}. The total number of injected
patterns is upper bounded by
\begin{equation}
N_{\mathrm{test}}
\leq
\sum_{a=2}^{5}
N_a
\sum_{w=1}^{4}
\binom{a}{w}
=
2\,687\,004\,851.
\label{eq:number-tested-patterns}
\end{equation}
This is an upper bound because the same fault support may occur in
multiple LETS instances.
For both the error-injection experiment and the Monte Carlo simulations,
we fix the initial decoder priors to those obtained at
\(p_0=10^{-3}\). Thus, the decoder configuration is kept fixed
throughout the physical-error-rate sweep, while the probability assigned
to each identified fault support is evaluated at the corresponding
physical error rate \(p\) through
Eq.~\eqref{eq:fault-support-probability}.

\subsection{Deterministic decoders}

For a deterministic decoder, every tested support has a fixed decoding
outcome. Define
\begin{equation*}
\mathcal{F}_w
=
\left\{
S\in\mathcal{U}_w:
\text{the decoder fails when \(S\) is injected}
\right\},
\label{eq:failing-supports}
\end{equation*}
with $w \in \{1,2,3,4\}$.
The sets \(\mathcal{F}_1\) and \(\mathcal{F}_2\) are empty for the
deterministic decoder considered here, since no failures were observed
at these weights. Since the sets \(\mathcal{F}_w\) contain only a
subset of all possible physical fault configurations, their total
probability gives a lower bound on the logical failure probability of
the complete \(r\)-round experiment:
\begin{equation}
P_{\mathrm{LER}}^{(r)}(p)
\geq
P_{\mathrm{LETS}}^{(r)}(p)
:=
\sum_{w=1}^{4}
\sum_{S\in\mathcal{F}_w}
\Pr(S;p).
\label{eq:deterministic-ef-estimate}
\end{equation}
For a deterministic decoder,
Eq.~\eqref{eq:deterministic-ef-estimate} is an exact evaluation of the
logical-error contribution associated with the identified
LETS-supported fault patterns of weights up to four. It remains a lower
bound on the complete logical failure probability because failures
outside the tested support universe are not included.

\subsection{Randomized decoders}
\label{sec:randomized-ef-estimation}

Both \texttt{RelayBP} and \texttt{ELMS} contain randomized components.
For these decoders, a fixed fault support \(S\) is characterized by the
conditional failure probability
\begin{equation*}
q_S
=
\Pr
\left(
\text{decoder failure when \(S\) is injected}
\right),
\label{eq:conditional-failure-probability}
\end{equation*}
The averaged
contribution of the tested LETS-supported patterns is therefore
\begin{equation}
\overline{P}_{\mathrm{LETS}}^{(r)}(p)
=
\sum_{w=1}^{4}
\sum_{S\in\mathcal{U}_w}
q_S\Pr(S;p),
\label{eq:randomized-lets-contribution}
\end{equation}
and remains a lower bound on the complete randomness-averaged logical
failure probability.
In the error-injection experiment, each distinct support is decoded
once. Let \(X_S\in\{0,1\}\) denote the resulting outcome, with
\(X_S=1\) if decoding fails. We estimate the LETS contribution as
\begin{equation}
\widehat{P}_{\mathrm{LETS}}^{(r)}(p)
=
\sum_{w=1}^{4}
\sum_{S\in\mathcal{U}_w}
X_S\Pr(S;p).
\label{eq:randomized-decoder-estimator}
\end{equation}
Since \(\mathbb{E}[X_S]=q_S\), it follows that
\begin{equation*}
\mathbb{E}
\left[
\widehat{P}_{\mathrm{LETS}}^{(r)}(p)
\right]
=
\overline{P}_{\mathrm{LETS}}^{(r)}(p).
\label{eq:randomized-decoder-unbiasedness}
\end{equation*}
Thus, using one execution per support introduces no systematic bias,
although it produces statistical fluctuations around the averaged LETS
contribution.
Assuming that the decoder randomness is sampled independently for
different injected supports, the variance of the estimator is
\begin{equation*}
\operatorname{Var}
\left[
\widehat{P}_{\mathrm{LETS}}^{(r)}(p)
\right]
=
\sum_{w=1}^{4}
\sum_{S\in\mathcal{U}_w}
q_S(1-q_S)\Pr(S;p)^2.
\label{eq:randomized-decoder-variance}
\end{equation*}
Since \(q_S(1-q_S)\leq 1/4\), this gives the decoder-independent bound
\begin{equation}
\operatorname{Var}
\left[
\widehat{P}_{\mathrm{LETS}}^{(r)}(p)
\right]
\leq
\frac{1}{4}
\sum_{w=1}^{4}
\sum_{S\in\mathcal{U}_w}
\Pr(S;p)^2.
\label{eq:randomized-decoder-variance-bound}
\end{equation}
The observed weight-one and weight-two supports have \(X_S=0\) for
both randomized decoders and therefore make no contribution to the
realized estimator in Eq.~\eqref{eq:randomized-decoder-estimator}.
Each execution already contains substantial internal diversity. In
particular, \texttt{ELMS} runs an ensemble of 20 layered min-sum
decoders with independently randomized layer orders and returns the
minimum-weight syndrome-valid estimate. Similarly, \texttt{RelayBP}
explores multiple randomized relay legs and selects the best valid
estimate found. Therefore, \(X_S\) represents the outcome of the
complete ensemble or multi-leg decoder, rather than that of a single
randomized message-passing realization.
A realization of
\(\widehat{P}_{\mathrm{LETS}}^{(r)}\) should not be interpreted as a
strict deterministic lower bound, since it may fluctuate around
\(\overline{P}_{\mathrm{LETS}}^{(r)}\). Rather, it is an unbiased
estimate of the lower-bound contribution associated with the tested
LETS-supported fault patterns. The close agreement between the
\texttt{RelayBP} estimate and the independent Monte Carlo simulation
indicates that these fluctuations are small at the aggregate level.

\subsection{Logical error rate per round}

We apply the preceding construction to the DEM of the
\([[144,12,12]]\) BB code with \(r=12\) rounds of syndrome extraction,
as implemented in
Refs.~\cite{bravyi2024high,gong2024lowlatencyiterativedecodingqldpc}.
The detector error model describes the complete memory experiment, so
the quantities above represent logical failure probabilities over all
\(r\) rounds.
For compactness, let \(P_{\mathrm{est}}^{(r)}(p)\) denote
\(P_{\mathrm{LETS}}^{(r)}(p)\) for a deterministic decoder and
\(\widehat{P}_{\mathrm{LETS}}^{(r)}(p)\) for a randomized decoder. We
convert this experiment-level probability to an effective logical error
rate per round according to
\begin{equation}
    P_{\mathrm{est,round}}(p)
    =
    1-
    \left[
        1-P_{\mathrm{est}}^{(r)}(p)
    \right]^{1/r}.
    \label{eq:ler-per-round}
\end{equation}
The same conversion is applied to the logical failure probabilities
obtained from the Monte Carlo simulations.

\section{Results}
\label{sec:results}

In this section, we compare the LETS-based error-floor estimates with
Monte Carlo simulations of \texttt{RelayBP}, \texttt{ImpulseBP}, and
\texttt{ELMS}, and analyze how the observed decoding failures are
distributed across LETS classes and topology classes. Each Monte Carlo
data point is obtained by observing at least \(100\) logical failures and
by performing at least \(10\,000\) decoding trials.

For each LETS class \((a,b)\), \(N_{\mathrm{TS}}\) denotes the number of
tested LETS instances, while \(N_{\mathrm{fail}}\) denotes the number of
instances for which at least one of the injected weight-three or
weight-four fault patterns causes a decoding failure. We define
\(f_{\mathrm{TS}}=100\times N_{\mathrm{fail}}/N_{\mathrm{TS}}\).
Furthermore, \(N_{\mathrm{top}}\) denotes the number of distinct
topology classes, where two LETSs belong to the same topology
class when their induced bipartite subgraphs are isomorphic.
The quantity \(N_{\mathrm{harm}}\) denotes the number of topology
classes containing at least one failing error pattern.

LETSs with \(a=2\) or \(a=3\) are omitted from the following tables
because none of their tested fault patterns causes a failure for any
of the three decoders. The reported classes with \(a=4\) or \(a=5\)
contain \(2851\) distinct LETS topology classes in total.

\subsection{Error-floor estimates and class-level statistics}

\begin{figure}
    \centering
    \includegraphics[width=\linewidth]{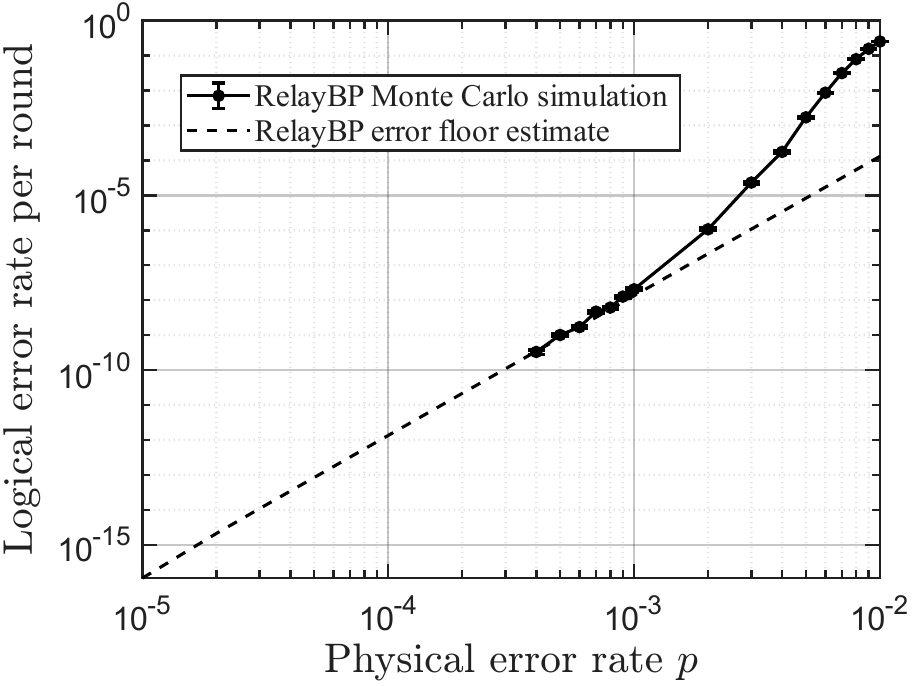}
    \caption{
        Comparison between the Monte Carlo simulation of
        \texttt{RelayBP} and the error-floor estimate obtained from
        weight-three and weight-four fault patterns supported on the
        enumerated LETSs. Both curves show the effective logical error
        rate per syndrome-extraction round.
    }
    \label{fig:ef_estimate_relaybp}
\end{figure}

Figure~\ref{fig:ef_estimate_relaybp} compares the Monte Carlo
simulation of \texttt{RelayBP} with the estimate obtained from the
enumerated LETSs. We use \(301\) relay legs and stop at the first
syndrome-valid estimate, corresponding to \textit{RelayBP1}
in Ref.~\cite{muller2025improvedbeliefpropagationsufficient}.
The two curves agree closely in the low-error-rate regime. Their
separation at larger physical error rates is expected because the
estimate retains only the identified weight-three and weight-four
failure mechanisms.

\begin{table}
    \centering
    \small
    \setlength{\tabcolsep}{7pt}
    \begin{tabular}{@{}c r r r r r@{}}
        \toprule
        Class
        & \(N_{\mathrm{TS}}\)
        & \(N_{\mathrm{fail}}\)
        & \(f_{\mathrm{TS}}\) (\%)
        & \(N_{\mathrm{top}}\)
        & \(N_{\mathrm{harm}}\) \\
        \midrule
        \((4,0)\) & 39\,995      & 0      & 0      & 16    & 0  \\
        \((4,2)\) & 176\,256     & 0      & 0      & 26    & 0  \\
        \((4,3)\) & 524\,088     & 0      & 0      & 63    & 0  \\
        \((4,4)\) & 1\,398\,780  & 23\,751 & 1.6980 & 85    & 1  \\
        \((4,5)\) & 2\,496\,600  & 1\,425  & 0.0571 & 119   & 4  \\
        \((5,0)\) & 194\,457     & 0      & 0      & 46    & 0  \\
        \((5,2)\) & 1\,257\,912  & 0      & 0      & 146   & 0  \\
        \((5,3)\) & 6\,824\,502  & 68\,239 & 0.9999 & 433   & 2  \\
        \((5,4)\) & 22\,244\,970 & 13\,045 & 0.0586 & 654   & 16 \\
        \((5,5)\) & 56\,669\,793 & 46\,397 & 0.0819 & 1\,263 & 48 \\
        \bottomrule
    \end{tabular}
    \caption{
        Class-level exhaustive error-injection results for
        \texttt{RelayBP}. An LETS instance is classified as failing if
        at least one tested weight-three or weight-four fault pattern
        supported on it causes a decoding failure.
        \(N_{\mathrm{top}}\) is the number of distinct induced-subgraph
        topology classes and \(N_{\mathrm{harm}}\) is the number of
        those classes containing at least one failing instance.
    }
        \label{tab:relay-ts-statistics}
\end{table}

Table~\ref{tab:relay-ts-statistics} reports the corresponding
class-level results. In total, \(152\,857\) LETS instances contain at
least one failing injected pattern. No failures are observed in the
\((4,0)\), \((4,2)\), \((4,3)\), \((5,0)\), or \((5,2)\) classes.
The largest failing-instance fractions occur for \((4,4)\) and
\((5,3)\), for which \(f_{\mathrm{TS}}=1.6980\%\) and
\(f_{\mathrm{TS}}=0.9999\%\), respectively. By contrast, the
fractions for \((4,5)\), \((5,4)\), and \((5,5)\) remain below
\(0.1\%\). The harmfulness of a LETS class therefore does not increase
monotonically with either \(a\) or \(b\).
The failures are also strongly concentrated at the topology level.
Only \(71\) of the \(2851\) topology classes, corresponding to
\(2.4904\%\), contain any failing \texttt{RelayBP} instance.
For example, although the \((5,3)\) class contains \(433\) distinct
topologies, only two of them are harmful.

\begin{figure}
    \centering
    \includegraphics[width=\linewidth]{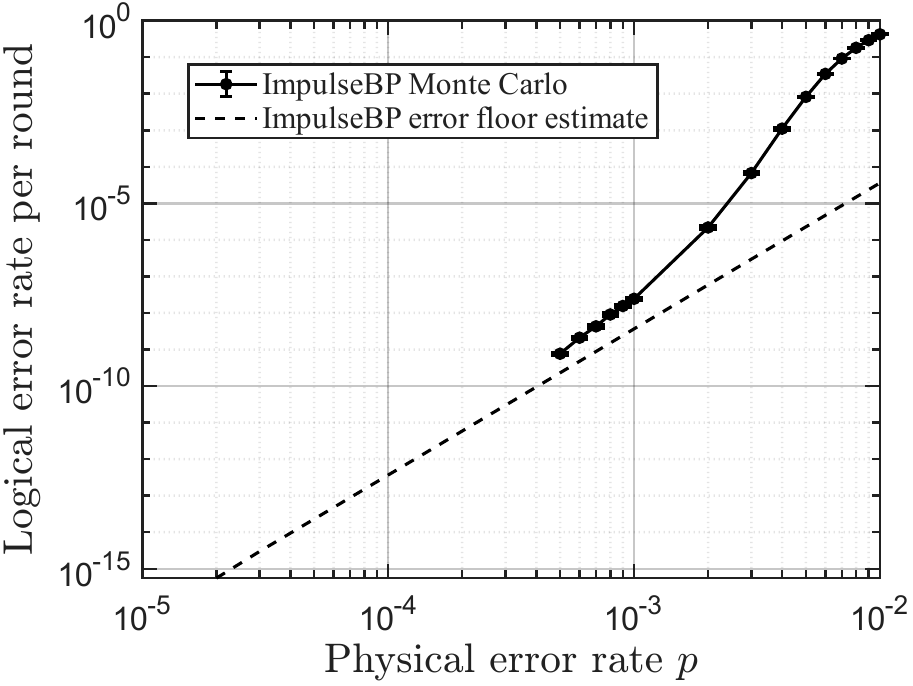}
    \caption{
        Comparison between the Monte Carlo simulation of
        \texttt{ImpulseBP} and the error-floor estimate obtained from
        weight-three and weight-four fault patterns supported on the
        enumerated LETSs. Both curves show the effective logical error
        rate per syndrome-extraction round.
    }
    \label{fig:impulse_ef}
\end{figure}

\begin{table}
    \centering
    \small
    \setlength{\tabcolsep}{7pt}
    \begin{tabular}{@{}c r r r r r@{}}
        \toprule
        Class
        & \(N_{\mathrm{TS}}\)
        & \(N_{\mathrm{fail}}\)
        & \(f_{\mathrm{TS}}\) (\%)
        & \(N_{\mathrm{top}}\)
        & \(N_{\mathrm{harm}}\) \\
        \midrule
        \((4,0)\) & 39\,995      & 0      & 0      & 16    & 0  \\
        \((4,2)\) & 176\,256     & 0      & 0      & 26    & 0  \\
        \((4,3)\) & 524\,088     & 0      & 0      & 63    & 0  \\
        \((4,4)\) & 1\,398\,780  & 5\,976  & 0.4272 & 85    & 1  \\
        \((4,5)\) & 2\,496\,600  & 576     & 0.0231 & 119   & 2  \\
        \((5,0)\) & 194\,457     & 0      & 0      & 46    & 0  \\
        \((5,2)\) & 1\,257\,912  & 0      & 0      & 146   & 0  \\
        \((5,3)\) & 6\,824\,502  & 18\,000 & 0.2638 & 433   & 1  \\
        \((5,4)\) & 22\,244\,970 & 4\,824  & 0.0217 & 654   & 9  \\
        \((5,5)\) & 56\,669\,793 & 13\,248 & 0.0234 & 1\,263 & 15 \\
        \bottomrule
    \end{tabular}
        \caption{
        Class-level exhaustive error-injection results for
        \texttt{ImpulseBP}. An LETS instance is classified as failing
        if at least one tested weight-three or weight-four fault pattern
        supported on it causes a decoding failure.
        \(N_{\mathrm{top}}\) is the number of distinct induced-subgraph
        topology classes and \(N_{\mathrm{harm}}\) is the number of
        those classes containing at least one failing instance.
    }
    \label{tab:impulse-ts-statistics}
\end{table}

Figure~\ref{fig:impulse_ef} compares the Monte Carlo simulation of
\texttt{ImpulseBP} with the corresponding LETS-based estimate. We use
\texttt{ImpulseBP}\((20,6)\), as defined in
Ref.~\cite{bhatnagar2026impulsedecodingquantumldpc}, with \(20\)
parallel constituent decoders, six residual reinitializations, and at
most \(100\) iterations for each constituent decoder. The estimate
lies below the simulated curve but remains within one order of
magnitude in the low-error-rate regime.

The class-level results in Table~\ref{tab:impulse-ts-statistics} follow
the same qualitative pattern as those obtained for \texttt{RelayBP}.
No failures occur in the \((4,0)\), \((4,2)\), \((4,3)\), \((5,0)\),
or \((5,2)\) classes. The largest failing-instance fractions again
occur for \((4,4)\) and \((5,3)\), with
\(f_{\mathrm{TS}}=0.4272\%\) and \(f_{\mathrm{TS}}=0.2638\%\),
respectively. For every class in which both decoders fail,
\texttt{ImpulseBP} has a smaller failing-instance fraction than
\texttt{RelayBP}, with reductions ranging from approximately a factor
of \(2.5\) to a factor of \(4\). However, since the error floor estimate is lower than the actual simulated curve, there exist failures of \texttt{ImpulseBP} that are not currently captured by our LETS analysis; it could be possible that, to fully approach the simulated curve, one needs to extend the search to structures with $a=6$, ETS with leaves or non-elementary TS.

The concentration at the topology level is even stronger than for
\texttt{RelayBP}. Only \(28\) of the \(2851\) topology classes,
corresponding to \(0.9821\%\), contain a failing error pattern. In particular, all \(18\,000\) failures
in the \((5,3)\) class belong to a single topology, and all \(5\,976\)
failures in the \((4,4)\) class also belong to a single topology.
These results demonstrate that the parameters \((a,b)\) alone are
insufficient to characterize harmfulness: within a given class,
failures can be confined to a very small subset of the possible
internal LETS structures.

\begin{figure}
    \centering
    \includegraphics[width=\linewidth]{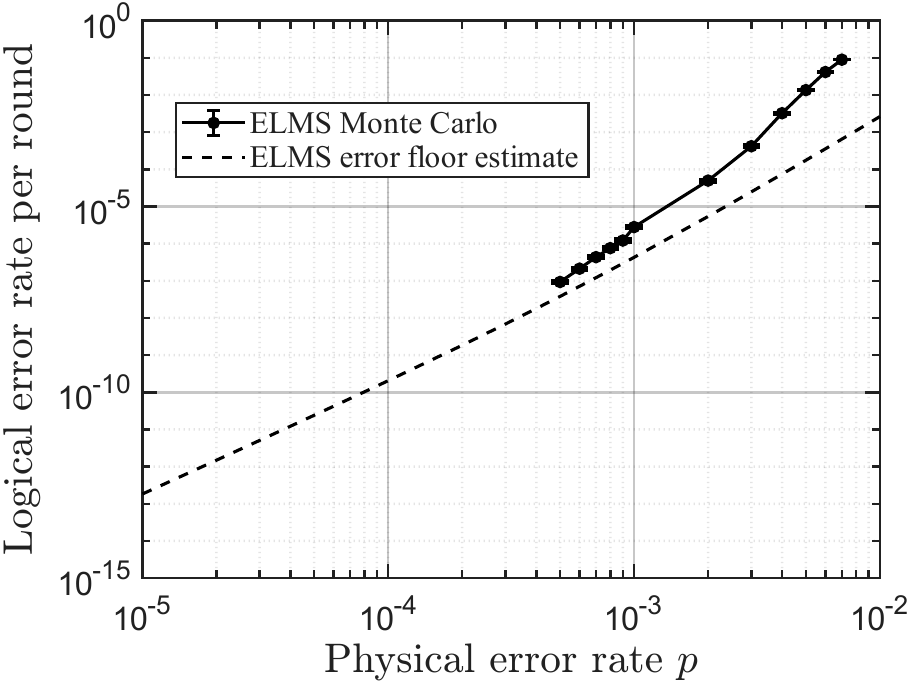}
    \caption{
        Comparison between the Monte Carlo simulation of
        \texttt{ELMS} and the error-floor estimate obtained from
        weight-three and weight-four fault patterns supported on the
        enumerated LETSs. Both curves show the effective logical error
        rate per syndrome-extraction round.
    }
    \label{fig:elms_ef}
\end{figure}

\begin{table}
    \centering
    \small
    \setlength{\tabcolsep}{7pt}
    \begin{tabular}{@{}c r r r r r@{}}
        \toprule
        Class
        & \(N_{\mathrm{TS}}\)
        & \(N_{\mathrm{fail}}\)
        & \(f_{\mathrm{TS}}\) (\%)
        & \(N_{\mathrm{top}}\)
        & \(N_{\mathrm{harm}}\) \\
        \midrule
        \((4,0)\) & 39\,995      & 0          & 0      & 16    & 0   \\
        \((4,2)\) & 176\,256     & 0          & 0      & 26    & 0   \\
        \((4,3)\) & 524\,088     & 3\,057      & 0.5833 & 63    & 5   \\
        \((4,4)\) & 1\,398\,780  & 79\,876     & 5.7104 & 85    & 11  \\
        \((4,5)\) & 2\,496\,600  & 87\,462     & 3.5032 & 119   & 30  \\
        \((5,0)\) & 194\,457     & 0          & 0      & 46    & 0   \\
        \((5,2)\) & 1\,257\,912  & 798        & 0.0634 & 146   & 2   \\
        \((5,3)\) & 6\,824\,502  & 388\,872    & 5.6982 & 433   & 35  \\
        \((5,4)\) & 22\,244\,970 & 847\,431    & 3.8095 & 654   & 165 \\
        \((5,5)\) & 56\,669\,793 & 1\,759\,046 & 3.1040 & 1\,263 & 399 \\
        \bottomrule
    \end{tabular}
     \caption{
        Class-level exhaustive error-injection results for
        \texttt{ELMS}. An LETS instance is classified as failing if at
        least one tested weight-three or weight-four fault pattern
        supported on it causes a decoding failure.
        \(N_{\mathrm{top}}\) is the number of distinct induced-subgraph
        topology classes and \(N_{\mathrm{harm}}\) is the number of
        those classes containing at least one failing instance.
    }
    \label{tab:elms-ts-statistics}
\end{table}

Figure~\ref{fig:elms_ef} compares the Monte Carlo simulation of
\texttt{ELMS} with its LETS-based estimate. As for
\texttt{ImpulseBP}, the estimate lies below the simulated curve but
remains within one order of magnitude in the low-error-rate regime.

Table~\ref{tab:elms-ts-statistics} shows that \texttt{ELMS} is more
susceptible to the enumerated LETSs than the other two decoders. In
total, \(3\,166\,542\) LETS instances contain at least one failing
pattern. No failures occur in the \((4,0)\), \((4,2)\), or \((5,0)\)
classes. The largest failing-instance fractions occur for \((4,4)\)
and \((5,3)\), with \(f_{\mathrm{TS}}=5.7104\%\) and
\(f_{\mathrm{TS}}=5.6982\%\), respectively. The \((4,5)\), \((5,4)\),
and \((5,5)\) classes also have substantial failing fractions between
approximately \(3.1\%\) and \(3.8\%\).

The \((5,2)\) class behaves differently: only \(798\) of its
\(1\,257\,912\) instances fail, giving
\(f_{\mathrm{TS}}=0.0634\%\). Although the \((5,5)\) class has the
largest absolute number of failing instances, its failing-instance
fraction is smaller than those of \((4,4)\), \((5,3)\), and \((5,4)\).
As for the other decoders, harmfulness is therefore not monotonic in
either \(a\) or \(b\).

The failures of \texttt{ELMS} are distributed across a broader
collection of structures. Of the \(2851\) topology classes, \(647\),
or \(22.6938\%\), contain at least one failing instance. This fraction
is substantially larger than the corresponding fractions for
\texttt{RelayBP} and \texttt{ImpulseBP}, coherently with the worse performance obtained through Monte Carlo simulation.

\begin{table*}
    \centering
    \small
    \setlength{\tabcolsep}{5pt}
    \begin{tabular}{@{}l c c c r r r r@{}}
        \toprule
        Decoder
        & Rank
        & Class
        & Topology
        & \(N_{\mathrm{TS}}(\tau)\)
        & \(N_{\mathrm{fail}}(\tau)\)
        & \(f_{\mathrm{TS}}(\tau)\) (\%)
        & \(s_{\mathrm{fail}}(\tau)\) (\%) \\
        \midrule
        \texttt{RelayBP}
        & 1 & \((5,3)\) & \(\tau_1\)
        & 90\,503 & 67\,912 & 75.0384 & 44.4285 \\
        \texttt{RelayBP}
        & 2 & \((5,5)\) & \(\tau_3\)
        & 60\,696 & 24\,946 & 41.0999 & 16.3198 \\
        \texttt{RelayBP}
        & 3 & \((4,4)\) & \(\tau_2\)
        & 489\,708 & 23\,751 & 4.8500 & 15.5381 \\
        \midrule
        \texttt{ImpulseBP}
        & 1 & \((5,3)\) & \(\tau_1\)
        & 90\,503 & 18\,000 & 19.8888 & 42.2297 \\
        \texttt{ImpulseBP}
        & 2 & \((4,4)\) & \(\tau_2\)
        & 489\,708 & 5\,976 & 1.2203 & 14.0203 \\
        \texttt{ImpulseBP}
        & 3 & \((5,5)\) & \(\tau_3\)
        & 60\,696 & 5\,040 & 8.3037 & 11.8243 \\
        \midrule
        \texttt{ELMS}
        & 1 & \((5,5)\) & \(\tau_4\)
        & 7\,151\,176 & 506\,182 & 7.0783 & 15.9853 \\
        \texttt{ELMS}
        & 2 & \((5,3)\) & \(\tau_5\)
        & 616\,890 & 264\,797 & 42.9245 & 8.3623 \\
        \texttt{ELMS}
        & 3 & \((5,4)\) & \(\tau_6\)
        & 3\,093\,242 & 208\,352 & 6.7357 & 6.5798 \\
        \bottomrule
    \end{tabular}
    \caption{
        Three highest-ranked LETS topology classes for each decoder,
        ordered by the number of failing LETS instances.
        \(N_{\mathrm{TS}}(\tau)\) and
        \(N_{\mathrm{fail}}(\tau)\) denote the total and failing
        numbers of instances belonging to topology class \(\tau\),
        respectively. The quantity \(f_{\mathrm{TS}}(\tau)\) is the
        percentage of instances of \(\tau\) that fail, while
        \(s_{\mathrm{fail}}(\tau)\) is its share of all failing LETS
        instances for the corresponding decoder. These are
        instance-level structural statistics and are not weighted by
        the physical probabilities of the injected fault supports.
    }
    \label{tab:top-ranked-ts-topologies}
\end{table*}

\subsection{Concentration and overlap of harmful topologies}

Table~\ref{tab:topology-concentration-summary} summarizes the
topology-level distribution of failures for the three decoders. A harmful
topology is classified as mixed if it contains both failing and
successfully decoded LETS, and as all-fail if all of its tested
LETSs fail for at least one error pattern. The final column gives the fraction of all failing
LETS instances belonging to the three highest-ranked topologies for
the corresponding decoder.

\begin{table}[t]
    \centering
    \small
    \setlength{\tabcolsep}{4pt}
    \begin{tabular}{@{}l r c r r@{}}
        \toprule
        Decoder
        & \(N_{\mathrm{fail}}^{\mathrm{tot}}\)
        & \(N_{\mathrm{harm}}\)
        & \(f_{\mathrm{harm}}\) (\%)
        & \(S_3\) (\%) \\
        \midrule
        \texttt{RelayBP}
        & 152\,857
        & 71 (69/2)
        & 2.4904
        & 76.2863 \\
        \texttt{ImpulseBP}
        & 42\,624
        & 28 (26/2)
        & 0.9821
        & 68.0743 \\
        \texttt{ELMS}
        & 3\,166\,542
        & 647 (638/9)
        & 22.6938
        & 30.9275 \\
        \bottomrule
    \end{tabular}
    \caption{
        Concentration of failing LETS instances across topology classes.
        Here, \(N_{\mathrm{fail}}^{\mathrm{tot}}\) is the total number of
        failing LETS instances, and \(N_{\mathrm{harm}}\) is the number
        of topology classes containing at least one failing instance.
        The values in parentheses give the numbers of mixed and all-fail
        topology classes, respectively.
        The fraction of harmful topologies is
        \(f_{\mathrm{harm}}=100N_{\mathrm{harm}}/2851\), where \(2851\)
        is the total number of topology classes in the analyzed range.
        Finally, \(S_3\) is the fraction of all failing instances
        belonging to the three highest-ranked topologies for each
        decoder.
    }
    \label{tab:topology-concentration-summary}
\end{table}

For both \texttt{RelayBP} and \texttt{ImpulseBP}, fewer than
\(2.5\%\) of the enumerated topology classes contain any failing
instance, and their three leading topologies account for
\(76.2863\%\) and \(68.0743\%\) of all failing instances,
respectively. Thus, the failures of these two decoders are highly
localized within the enumerated LETS space. The concentration is
weaker for \texttt{ELMS}: its three leading topologies account for
\(30.9275\%\) of the failing instances, consistent with its much larger
number of harmful topology classes.

Most harmful topologies are mixed rather than all-fail. In particular,
\(69\) of the \(71\) harmful \texttt{RelayBP} topologies, \(26\) of
the \(28\) harmful \texttt{ImpulseBP} topologies, and \(638\) of the
\(647\) harmful \texttt{ELMS} topologies contain both failing and
successfully decoded instances. The induced LETS topology is therefore
a strong indicator of decoder vulnerability, but it does not by itself
determine the decoding outcome. The specific location of the LETSs in the Tanner graph, its associated priors, and the particular injected pattern are all factors that influence the decoder dynamics, and that explain why different LETS in the same topology class may behave differently.

Table~\ref{tab:top-ranked-ts-topologies} gives the three highest-ranked
topologies for each decoder, based on \(N_{\mathrm{fail}}(\tau)\), the number of failing LETS instances
belonging to \(\tau\). For a topology class \(\tau\), we define
\(s_{\mathrm{fail}}(\tau)
=100N_{\mathrm{fail}}(\tau)/N_{\mathrm{fail}}^{\mathrm{tot}}\).

For ease of reference, we assign symbolic labels to the topology
classes discussed below. The three topology classes forming the
top-three set for both \texttt{RelayBP} and \texttt{ImpulseBP} are
denoted by \(\tau_1\), \(\tau_2\), and \(\tau_3\), corresponding to the
\((5,3)\), \((4,4)\), and \((5,5)\) classes, respectively. The three
additional topology classes forming the top-three set for
\texttt{ELMS} are denoted by \(\tau_4\), \(\tau_5\), and \(\tau_6\),
in decreasing order of their number of failing \texttt{ELMS}
instances.
The same three topology classes form the complete top-three set for
both \texttt{RelayBP} and \texttt{ImpulseBP}, although their ordering
differs. Topology \(\tau_1\) is ranked first for both decoders and
alone accounts for \(44.4285\%\) of the failing \texttt{RelayBP}
instances and \(42.2297\%\) of the failing \texttt{ImpulseBP}
instances. For \texttt{ImpulseBP}, all \(18\,000\) failing instances
in the entire \((5,3)\) class belong to \(\tau_1\). Similarly,
\(\tau_2\) contains all \(5\,976\) \texttt{ImpulseBP} failures and all
\(23\,751\) \texttt{RelayBP} failures in the \((4,4)\) class.
The concentration cannot be explained solely by the abundance of the
topology classes. For example, \(\tau_3\) represents only
\(60\,696/56\,669\,793=0.1071\%\) of the enumerated \((5,5)\)
instances, but accounts for \(24\,946/46\,397=53.7664\%\) of the
\texttt{RelayBP} failures and \(5\,040/13\,248=38.0435\%\) of the
\texttt{ImpulseBP} failures in that class. It follows that within a generic $(a,b)$ class, only specific topology classes are really harmful.
\begin{figure*}[t]
    \centering

    \begin{subfigure}[t]{0.32\textwidth}
        \centering
        \includegraphics[
            width=\linewidth
        ]{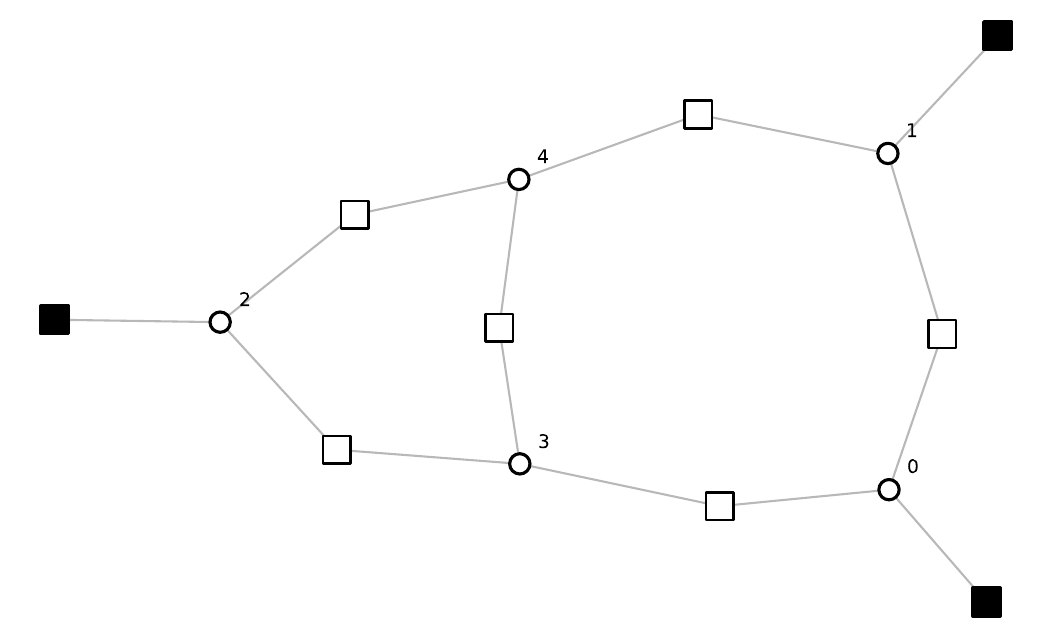}
        \caption{
            Topology \(\tau_1\) in the \((5,3)\) class.
        }
        \label{fig:shared_structure_5_3}
    \end{subfigure}
    \hfill
    \begin{subfigure}[t]{0.20\textwidth}
        \centering
        \includegraphics[
            width=\linewidth
        ]{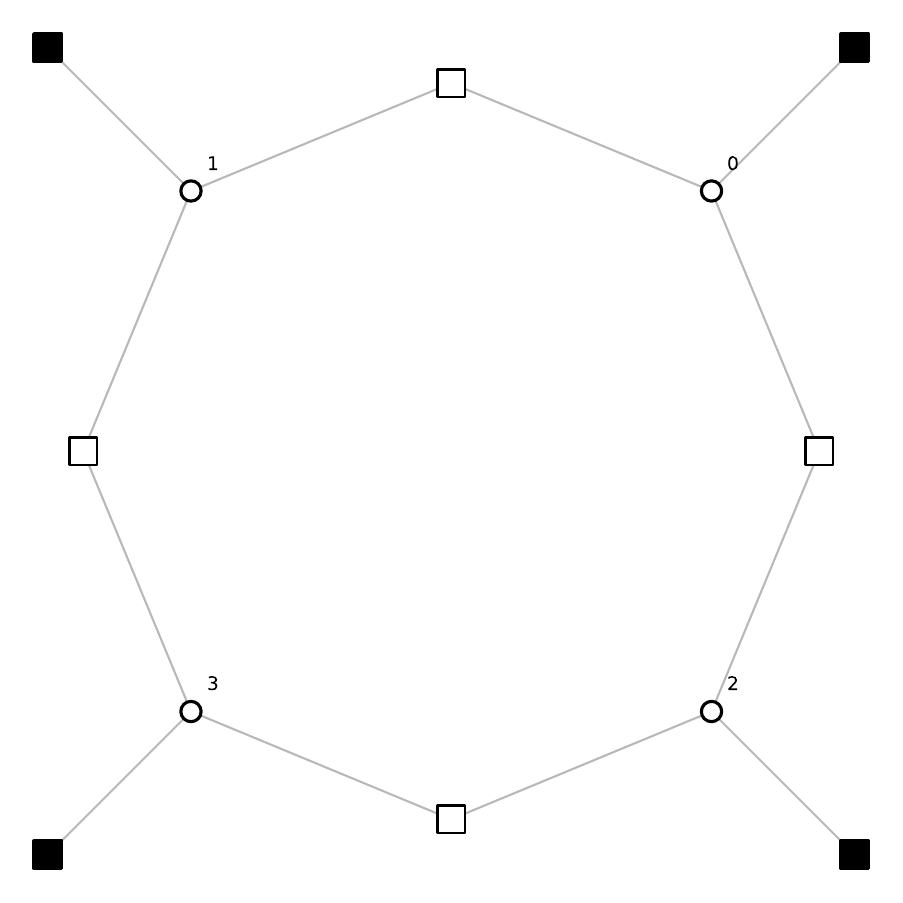}
        \caption{
            Topology \(\tau_2\) in the \((4,4)\) class.
        }
        \label{fig:shared_structure_4_4}
    \end{subfigure}
    \hfill
    \begin{subfigure}[t]{0.32\textwidth}
        \centering
        \includegraphics[
            width=\linewidth
        ]{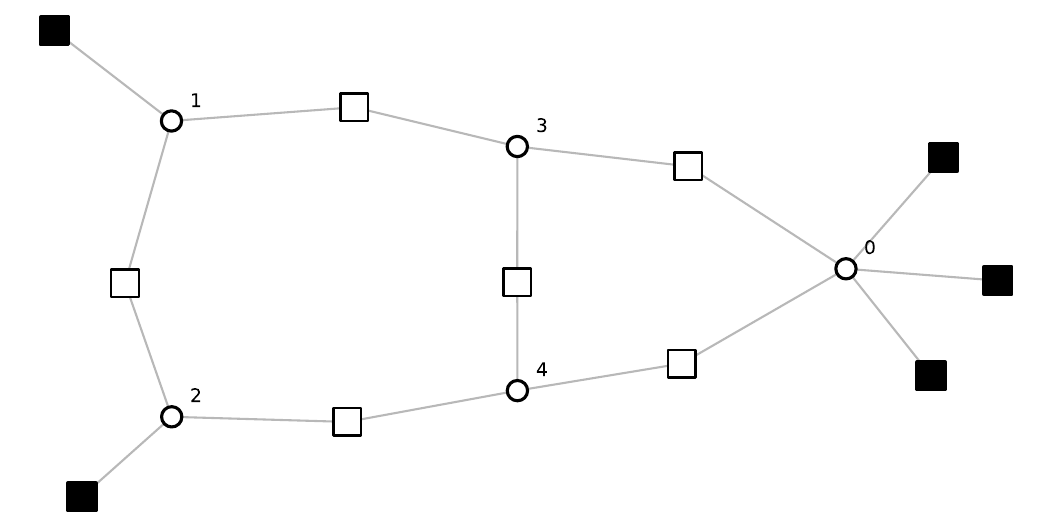}
        \caption{
            Topology \(\tau_3\) in the \((5,5)\) class.
        }
        \label{fig:shared_structure_5_5}
    \end{subfigure}

    \caption{
        Canonical representatives of the three LETS topology classes
        forming the top-three set for both \texttt{RelayBP} and
        \texttt{ImpulseBP}. The same structures remain highly ranked
        for \texttt{ELMS}: \(\tau_1\), \(\tau_2\), and \(\tau_3\)
        have ranks \(6\), \(10\), and \(17\), respectively, among its
        \(647\) harmful topology classes. Circles represent variable
        nodes and squares represent check nodes; white and black squares
        indicate satisfied and unsatisfied checks, respectively.
    }
    \label{fig:representative_structures}
\end{figure*}
The overlap is not restricted to \texttt{RelayBP} and
\texttt{ImpulseBP}. Table~\ref{tab:shared-topology-ranks} reports the
positions of \(\tau_1\), \(\tau_2\), and \(\tau_3\) for all three
decoders. Each decoder entry gives the rank among its harmful topology
classes, followed by the fraction of instances of that topology class
that fail.

\begin{table}[t]
    \centering
    \small
    \setlength{\tabcolsep}{6pt}
    \begin{tabular}{@{}c c c c c@{}}
        \toprule
        Topology
        & Class
        & \texttt{RelayBP}
        & \texttt{ImpulseBP}
        & \texttt{ELMS} \\
        \midrule
        \(\tau_1\)
        & \((5,3)\)
        & \(1/75.0384\%\)
        & \(1/19.8888\%\)
        & \(6/89.1650\%\) \\
        \(\tau_2\)
        & \((4,4)\)
        & \(3/4.8500\%\)
        & \(2/1.2203\%\)
        & \(10/12.2208\%\) \\
        \(\tau_3\)
        & \((5,5)\)
        & \(2/41.0999\%\)
        & \(3/8.3037\%\)
        & \(17/57.0664\%\) \\
        \bottomrule
    \end{tabular}
    \caption{
        Cross-decoder ranks of the three topology classes forming the
        top-three set for both \texttt{RelayBP} and
        \texttt{ImpulseBP}. Each decoder entry has the form
        ``rank/\(f_{\mathrm{TS}}(\tau)\)'', where the rank is among
        the harmful topology classes of that decoder and
        \(f_{\mathrm{TS}}(\tau)\) is the percentage of instances of
        topology class \(\tau\) that fail.
    }
    \label{tab:shared-topology-ranks}
\end{table}

\begin{figure*}[t]
    \centering

    \begin{subfigure}[t]{0.25\textwidth}
        \centering
        \includegraphics[
            width=\linewidth
        ]{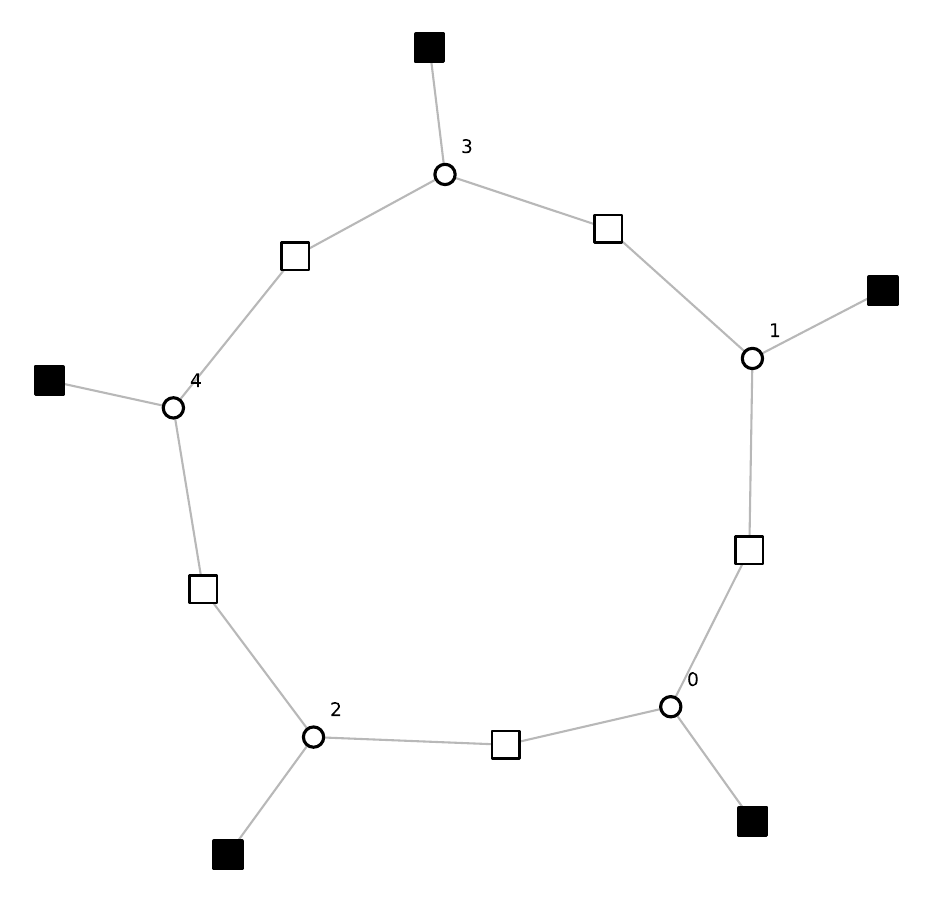}
        \caption{
            Topology \(\tau_4\) in the \((5,5)\) class.
        }
        \label{fig:elms_structure_rank_1}
    \end{subfigure}
    \hfill
    \begin{subfigure}[t]{0.32\textwidth}
        \centering
        \includegraphics[
            width=\linewidth
        ]{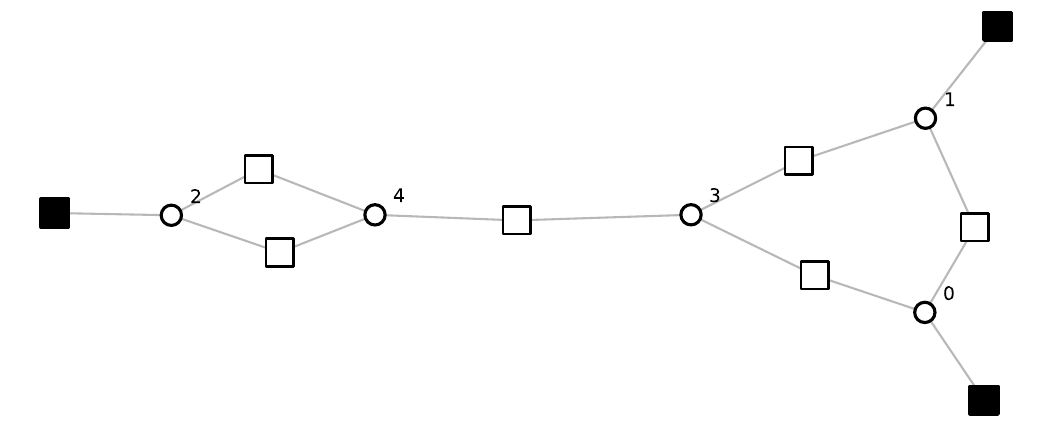}
        \caption{
            Topology \(\tau_5\) in the \((5,3)\) class.
        }
        \label{fig:elms_structure_rank_2}
    \end{subfigure}
    \hfill
    \begin{subfigure}[t]{0.25\textwidth}
        \centering
        \includegraphics[
            width=\linewidth
        ]{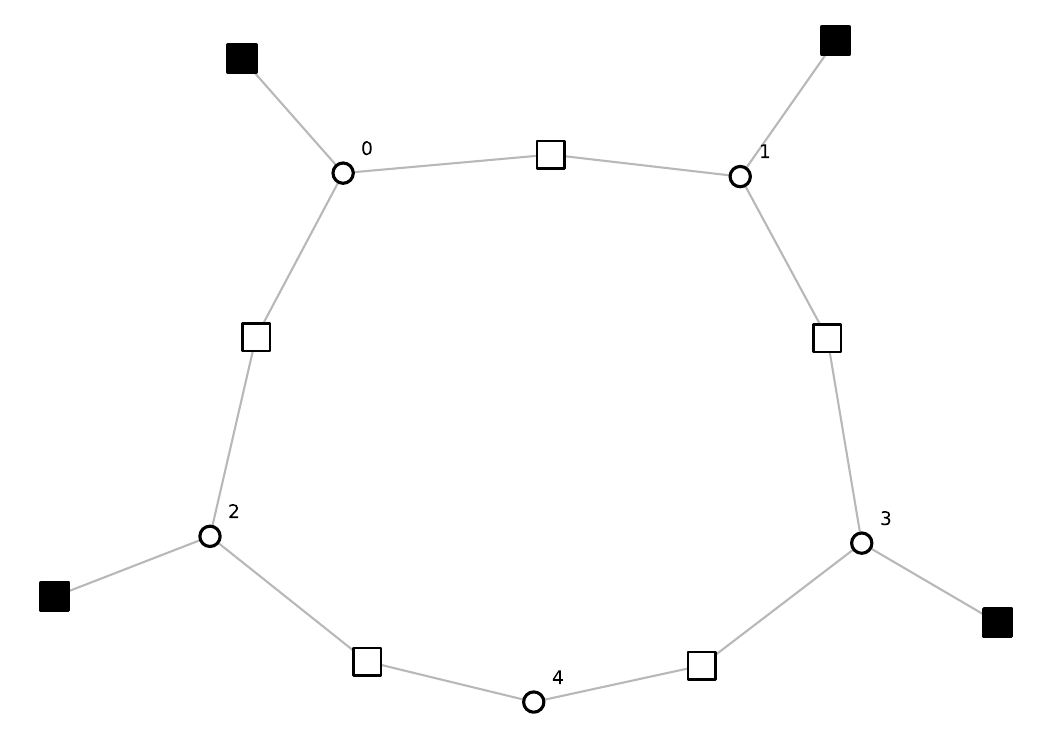}
        \caption{
            Topology \(\tau_6\) in the \((5,4)\) class.
        }
        \label{fig:elms_structure_rank_3}
    \end{subfigure}

    \caption{
        Canonical representatives of the three topology classes
        associated with the largest numbers of failing
        \texttt{ELMS} instances. The classes \(\tau_4\), \(\tau_5\),
        and \(\tau_6\) account for \(15.9853\%\), \(8.3623\%\), and
        \(6.5798\%\) of all failing \texttt{ELMS} instances,
        respectively, and for \(30.9275\%\) in total. Circles represent
        variable nodes and squares represent check nodes; white and
        black squares indicate satisfied and unsatisfied checks,
        respectively.
    }
    \label{fig:representative_structures_elms}
\end{figure*}

Although the leading \texttt{ELMS} topology classes differ from those
of the other two decoders, \(\tau_1\), \(\tau_2\), and \(\tau_3\)
remain among its \(17\) highest-ranked harmful topology classes.
Moreover, their failing-instance fractions are substantial:
\(89.1650\%\) for \(\tau_1\), \(12.2208\%\) for \(\tau_2\), and
\(57.0664\%\) for \(\tau_3\). Together, these three topology classes
account for \(5.5322\%\) of all failing \texttt{ELMS} instances. The
three decoders therefore do not have identical failure landscapes,
but they exhibit a common core of highly vulnerable local structures
despite their substantially different decoding heuristics.

The combined results establish two distinct features of the decoder
failure mechanisms. First, failures are highly nonuniform within each
\((a,b)\) class and are concentrated in particular internal LETS
topologies. Second, the topology classes \(\tau_1\), \(\tau_2\), and
\(\tau_3\) remain highly harmful across all three decoders, even though
the overall degree of concentration depends strongly on the decoder.
This structural overlap supports the use of the enumerated LETSs not
only for estimating low-error-rate performance, but also for
identifying localized graph configurations that can be targeted in the
design of improved iterative decoders.


\section{Conclusions}
\label{sec:conclusions}

In this paper, we introduced a framework for estimating the
error-floor performance of message-passing decoders for QLDPC codes
under circuit-level noise. Building on the classical trapping-set
literature, we applied the \texttt{dpl-search} algorithm directly to
the circuit-level DEM of the \( [[144,12,12]] \) BB code and enumerated
all LETSs in classes up to \((5,5)\). We then exhaustively injected
weight-three and weight-four fault patterns supported on the
enumerated LETSs and evaluated three circuit-level decoders with
substantially different architectures. The resulting estimate closely
agrees with the Monte Carlo simulation for \texttt{RelayBP}, while for
\texttt{ImpulseBP} and \texttt{ELMS} it remains within the same order
of magnitude.

Our results lead to several conclusions. First, exhaustive
trapping-set search algorithms can be applied directly to
circuit-level DEMs. However, the small girth and highly irregular
degree distributions of these graphs make the search computationally
prohibitive for \(a>5\) in the implementation considered here.
Although the resulting search depth appears sufficient for the
\( [[144,12,12]] \) code, it remains unclear whether the same approach
will remain both sufficient and computationally feasible for
larger-distance codes, such as the \( [[288,12,18]] \) BB code.

Second, restricting the analysis to small LETSs is sufficient to
reproduce the error floor of \texttt{RelayBP} and to recover a
substantial part of the low-weight contribution for
\texttt{ImpulseBP} and \texttt{ELMS}. This shows that trapping-set
analysis remains useful even for decoders that employ mechanisms such
as re-initialization, randomization, ensembles, and global processing,
for which a local structural characterization was not evident
\textit{a priori}. The identified harmful LETSs can therefore be used
not only to estimate error-floor performance, but also to guide the
design of improved decoding algorithms.

At the same time, the vast majority of the enumerated LETSs are not
associated with any observed decoding failure. Therefore, membership
in a given \((a,b)\) class does not by itself determine whether a LETS
is harmful. The decoding outcome also depends on its internal
topology, its location within the Tanner graph, the injected fault pattern, the
fault priors, and the specific decoder dynamics. Quantum degeneracy
and the presence of nearby or containing stabilizer structures may
also affect this behavior.

Third, the three highest-ranked topology classes are the same for
\texttt{RelayBP} and \texttt{ImpulseBP}, although their ordering is
different. These topology classes also remain among the most harmful
ones for \texttt{ELMS}, despite the broader distribution of its
failures. The three decoders therefore do not have identical failure
mechanisms, but they share a common set of particularly vulnerable
local structures. This suggests that a topology-level analysis is
more informative for decoder design than a classification based only
on the \((a,b)\) parameters.

These results motivate several directions for future work. Since most
of the enumerated LETSs are harmless for the decoders considered, the
search could be restricted to narrower structural subclasses that are
more strongly associated with decoding failure. This could make it
possible to explore larger \((a,b)\) classes without enumerating the
entire LETS space. A second direction is to exploit the catalog of
harmful LETSs identified in this work to design decoders that
explicitly suppress or escape these failure mechanisms. In
particular, all the analyzed decoders fail on some weight-three and
weight-four fault patterns. Eliminating these low-weight failures
would improve their error-floor performance, even without attaining
the full bounded-distance correction capability of the code, namely
the correction of all fault patterns of weight at most five.

\bibliography{referencesNEW}

@article{chytas_enhanced_2024,
author={Chytas, Dimitris and Pacenti, Michele and Raveendran, Nithin and Flanagan, Mark F. and Vasić, Bane},
  journal={IEEE Communications Letters}, 
  title={{Enhanced Message-Passing Decoding of Degenerate Quantum Codes Utilizing Trapping Set Dynamics}}, 
  year={2024},
  volume={28},
  number={3},
  pages={444-448},
  doi={10.1109/LCOMM.2024.3356312}}

@article{calderbank_good_1996,
	title = {{Good Quantum Error-Correcting Codes Exist}},
	volume = {54},
	number = {2},
	urldate = {2024-01-10},
	journal = {Physical Review A},
	author = {Calderbank, A. R. and Shor, Peter W.},
	month = aug,
	year = {1996},
	pages = {1098--1105},
    doi = {https://doi.org/10.1103/PhysRevA.54.1098}
}

@article{panteleev_degenerate_2021,
	title = {Degenerate {Quantum} {LDPC} {Codes} {With} {Good} {Finite} {Length} {Performance}},
	volume = {5},
	doi = {10.22331/q-2021-11-22-585},
	urldate = {2024-01-03},
	journal = {Quantum},
	author = {Panteleev, Pavel and Kalachev, Gleb},
	month = nov,
	year = {2021},
	pages = {585},
}

@article{raveendran_trapping_2021,
	title = {Trapping {Sets} of {Quantum} {LDPC} {Codes}},
	volume = {5},
	urldate = {2023-06-21},
	journal = {Quantum},
	author = {Raveendran, Nithin and Vasić, Bane},
	month = oct,
	year = {2021},
	pages = {562},
    doi = {	https://doi.org/10.22331/q-2021-10-14-562},
}

@article{bravyi2024high,
  title={High-{T}hreshold and {L}ow-{O}verhead {F}ault-{T}olerant {Q}uantum {M}emory},
  author={Bravyi, Sergey and Cross, Andrew W and Gambetta, Jay M and Maslov, Dmitri and Rall, Patrick and Yoder, Theodore J},
  journal={Nature},
  volume={627},
  number={8005},
  pages={778--782},
  year={2024},
  publisher={Nature Publishing Group UK London},
  doi = {https://doi.org/10.1038/s41586-024-07107-7}
}

@article{chytas2024collectivebitflippingbaseddecoding,
  author={Chytas, Dimitris and Raveendran, Nithin and Vasić, Bane},
  journal={IEEE Transactions on Communications}, 
  title={{Collective Bit Flipping-Based Decoding of Quantum LDPC Codes}}, 
  year={2025},
  volume={73},
  number={8},
  pages={5566-5579},
  doi={10.1109/TCOMM.2025.3535897},
  }

@inproceedings{chytas2025enhancedminsumdecodingquantum,
author={Chytas, Dimitris and Raveendran, Nithin and Vasić, Bane},
  booktitle={2025 IEEE International Symposium on Information Theory (ISIT)}, 
  title={{Enhanced Min-Sum Decoding of Quantum Codes with Iteration Dynamics Memory}}, 
  year={2025},
  volume={},
  number={},
  pages={1-6},
  doi={10.1109/ISIT63088.2025.11195509}}

@misc{gong2024lowlatencyiterativedecodingqldpc,
      title={{Toward Low-latency Iterative Decoding of QLDPC Codes Under Circuit-Level Noise}}, 
      author={Anqi Gong and Sebastian Cammerer and Joseph M. Renes},
      year={2024},
      eprint={2403.18901},
      archivePrefix={arXiv},
}

@article{fault_complex,
  title = {Single-shot and measurement-based quantum error correction via fault complexes},
  author = {Hillmann, Timo and Dauphinais, Guillaume and Tzitrin, Ilan and Vasmer, Michael},
  journal = {Phys. Rev. A},
  volume = {112},
  issue = {4},
  pages = {L040401},
  numpages = {7},
  year = {2025},
  month = {Oct},
  publisher = {American Physical Society},
  doi = {10.1103/cjb4-l57n},
  url = {https://link.aps.org/doi/10.1103/cjb4-l57n}
}

@misc{valentini2025restartbeliefgeneralquantum,
      title={{Restart Belief: A General Quantum LDPC Decoder}}, 
      author={Lorenzo Valentini and Diego Forlivesi and Andrea Talarico and Marco Chiani},
      year={2025},
      eprint={2511.13281},
      archivePrefix={arXiv},
}

@misc{muller2025improvedbeliefpropagationsufficient,
      title={Improved belief propagation is sufficient for real-time decoding of quantum memory}, 
      author={Tristan Müller and Thomas Alexander and Michael E. Beverland and Markus Bühler and Blake R. Johnson and Thilo Maurer and Drew Vandeth},
      year={2025},
      eprint={2506.01779},
      archivePrefix={arXiv},

}

@article{ye2025beamsearchdecoderquantum,
  title = {Beam Search Decoder for Quantum Low-Density Parity-Check Codes},
  author = {Ye, Min and Wecker, Dave and Delfosse, Nicolas},
  journal = {PRX Quantum},
  volume = {7},
  issue = {3},
  pages = {033002},
  numpages = {12},
  year = {2026},
  month = {Jul},
  publisher = {American Physical Society},
  doi = {10.1103/6k5x-ztqt},
  url = {https://link.aps.org/doi/10.1103/6k5x-ztqt}
}

@ARTICLE{richardson-stopping-sets,
  author={Changyan Di and Proietti, D. and Telatar, I.E. and Richardson, T.J. and Urbanke, R.L.},
  journal={IEEE Transactions on Information Theory}, 
  title={Finite-length analysis of low-density parity-check codes on the binary erasure channel}, 
  year={2002},
  volume={48},
  number={6},
  pages={1570-1579},
  doi={10.1109/TIT.2002.1003839}}

@ARTICLE{dolecek_as,
  author={Amiri, Behzad and Lin, Chi-Wei and Dolecek, Lara},
  journal={IEEE Transactions on Communications}, 
  title={Asymptotic Distribution of Absorbing Sets and Fully Absorbing Sets for Regular Sparse Code Ensembles}, 
  year={2013},
  volume={61},
  number={2},
  pages={455-464},
  doi={10.1109/TCOMM.2012.120512.110605}}

@misc{morris2024absorbingsetsquantumldpc,
      title={{Absorbing Sets in Quantum LDPC Codes}}, 
      author={Kirsten D. Morris and Tefjol Pllaha and Christine A. Kelley},
      year={2024},
      eprint={2307.14532},
      archivePrefix={arXiv},}

@INPROCEEDINGS{ontology,
  author={Vasić, Bane and Chilappagari, Shashi Kiran and Nguyen, Dung Viet and Planjery, Shiva Kumar},
  booktitle={2009 47th Annual Allerton Conference on Communication, Control, and Computing (Allerton)}, 
  title={Trapping set ontology}, 
  year={2009},
  volume={},
  number={},
  pages={1-7},
  doi={10.1109/ALLERTON.2009.5394825}}

@ARTICLE{6192273,
  author={Kyung, Gyu Bum and Wang, Chih-Chun},
  journal={IEEE Transactions on Communications}, 
  title={Finding the Exhaustive List of Small Fully Absorbing Sets and Designing the Corresponding Low Error-Floor Decoder}, 
  year={2012},
  volume={60},
  number={6},
  pages={1487-1498},
  doi={10.1109/TCOMM.2012.042712.100672}}

@ARTICLE{bani_regular,
  author={Hashemi, Yoones and Banihashemi, Amir H.},
  journal={IEEE Transactions on Information Theory}, 
  title={{New Characterization and Efficient Exhaustive Search Algorithm for Leafless Elementary Trapping Sets of Variable-Regular LDPC Codes}}, 
  year={2016},
  volume={62},
  number={12},
  pages={6713-6736},
  doi={10.1109/TIT.2016.2613113}}

@ARTICLE{bani_irregular,
  author={Hashemi, Yoones and Banihashemi, Amir H.},
  journal={IEEE Transactions on Information Theory}, 
  title={{Characterization of Elementary Trapping Sets in Irregular LDPC Codes and the Corresponding Efficient Exhaustive Search Algorithms}}, 
  year={2018},
  volume={64},
  number={5},
  pages={3411-3430},
  doi={10.1109/TIT.2018.2799627}}

@article{Maan2026,
  author  = {Maan, Arshpreet Singh and Garcia Herrero, Francisco Miguel and Paler, Alexandru and Savin, Valentin},
  title   = {Decoding correlated errors in quantum {LDPC} codes},
  journal = {Nature Communications},
  year    = {2026},
  volume  = {17},
  number  = {1},
  pages   = {3965},
  doi     = {10.1038/s41467-026-70556-3},
  url     = {https://doi.org/10.1038/s41467-026-70556-3},
  issn    = {2041-1723}
}

@misc{bhatnagar2026impulsedecodingquantumldpc,
      title={Impulse Decoding of Quantum LDPC Codes: Equivalence of Degeneracy and Code-Shortening}, 
      author={Shobhit Bhatnagar and Michele Pacenti and Nithin Raveendran and David Declercq and Bane Vasić},
      year={2026},
      eprint={2606.18240},
      archivePrefix={arXiv}, 
}

@article{Battaglioni2023,
  author  = {Battaglioni, Massimo and Chiaraluce, Franco and Baldi, Marco and Pacenti, Michele and Mitchell, David G. M.},
  title   = {Optimizing quasi-cyclic spatially coupled {LDPC} codes by eliminating harmful objects},
  journal = {EURASIP Journal on Wireless Communications and Networking},
  year    = {2023},
  volume  = {2023},
  number  = {1},
  pages   = {67},
  doi     = {10.1186/s13638-023-02273-0},
  url     = {https://doi.org/10.1186/s13638-023-02273-0},
  issn    = {1687-1499}
}

@ARTICLE{faid1,
  author={Planjery, Shiva Kumar and Declercq, David and Danjean, Ludovic and Vasic, Bane},
  journal={IEEE Transactions on Communications}, 
  title={{Finite Alphabet Iterative Decoders—Part I: Decoding Beyond Belief Propagation on the Binary Symmetric Channel}}, 
  year={2013},
  volume={61},
  number={10},
  pages={4033-4045},
  doi={10.1109/TCOMM.2013.090513.120443}}

@ARTICLE{faid2,
  author={Declercq, David and Vasic, Bane and Planjery, Shiva Kumar and Li, Erbao},
  journal={IEEE Transactions on Communications}, 
  title={{Finite Alphabet Iterative Decoders—Part II: Towards Guaranteed Error Correction of LDPC Codes via Iterative Decoder Diversity}}, 
  year={2013},
  volume={61},
  number={10},
  pages={4046-4057},
  doi={10.1109/TCOMM.2013.090513.120444}}

@ARTICLE{6567866,
  author={Zhang, Shuai and Schlegel, Christian},
  journal={IEEE Transactions on Communications}, 
  title={{Controlling the Error Floor in LDPC Decoding}}, 
  year={2013},
  volume={61},
  number={9},
  pages={3566-3575},
  doi={10.1109/TCOMM.2013.071813.120659}}

@ARTICLE{6169192,
  author={Nguyen, Dung Viet and Chilappagari, Shashi Kiran and Marcellin, Michael W. and Vasic, Bane},
  journal={IEEE Transactions on Information Theory}, 
  title={On the Construction of Structured LDPC Codes Free of Small Trapping Sets}, 
  year={2012},
  volume={58},
  number={4},
  pages={2280-2302},
  doi={10.1109/TIT.2011.2173733}}

@INPROCEEDINGS{6034068,
  author={Wang, Jiadong and Dolecek, Lara and Zhang, Zhengya and Wesel, Richard},
  booktitle={2011 IEEE International Symposium on Information Theory Proceedings}, 
  title={Absorbing set spectrum approach for practical code design}, 
  year={2011},
  volume={},
  number={},
  pages={2726-2730},
  doi={10.1109/ISIT.2011.6034068}}

@INPROCEEDINGS{asit_istc,
  author={Pradhan, Asit Kumar and Raveendran, Nithin and Rengaswamy, Narayanan and Xiao, Xin and Vasić, Bane},
  booktitle={2023 12th International Symposium on Topics in Coding (ISTC)}, 
  title={Learning to Decode Trapping Sets in QLDPC Codes}, 
  year={2023},
  volume={},
  number={},
  pages={1-5},
  doi={10.1109/ISTC57237.2023.10273526}}

@ARTICLE{asit_tinfo,
  author={Pradhan, Asit Kumar and Raveendran, Nithin and Rengaswamy, Narayanan and Vasić, Bane},
  journal={IEEE Transactions on Information Theory}, 
  title={Linear Time Iterative Decoders for Hypergraph-Product and Lifted-Product Codes}, 
  year={2026},
  volume={72},
  number={7},
  pages={4618-4647},
  doi={10.1109/TIT.2026.3687799}}

@misc{beverland2025failfasttechniquesprobe,
      title={Fail fast: techniques to probe rare events in quantum error correction}, 
      author={Michael E. Beverland and Malcolm Carroll and Andrew W. Cross and Theodore J. Yoder},
      year={2025},
      eprint={2511.15177},
      archivePrefix={arXiv}, 
}

@ARTICLE{raveendran_expansion,
  author={Raveendran, Nithin and Declercq, David and Vasić, Bane},
  journal={IEEE Transactions on Communications}, 
  title={A Sub-Graph Expansion-Contraction Method for Error Floor Computation}, 
  year={2020},
  volume={68},
  number={7},
  pages={3984-3995},
  doi={10.1109/TCOMM.2020.2988676}}

@misc{yin2024symbreakmitigatingquantumdegeneracy,
      title={{SymBreak: Mitigating Quantum Degeneracy Issues in QLDPC Code Decoders by Breaking Symmetry}}, 
      author={Keyi Yin and Xiang Fang and Jixuan Ruan and Hezi Zhang and Dean Tullsen and Andrew Sornborger and Chenxu Liu and Ang Li and Travis Humble and Yufei Ding},
      year={2024},
      eprint={2412.02885},
      archivePrefix={arXiv},
}

@misc{tsubouchi2026degeneracycuttinglocalefficient,
      title={{Degeneracy Cutting: A Local and Efficient Post-Processing for Belief Propagation Decoding of Quantum Low-Density Parity-Check Codes}}, 
      author={Kento Tsubouchi and Hayata Yamasaki and Shiro Tamiya},
      year={2026},
      eprint={2510.08695},
      archivePrefix={arXiv},
}

@INPROCEEDINGS{pacenti2025turbo,
  author={Pacenti, Michele and Pradhan, Asit K. and Borah, Shantom K. and Vasić, Bane},
  booktitle={2025 IEEE International Conference on Quantum Computing and Engineering (QCE)}, 
  title={{Turbo-Annihilation of Hook Errors in Stabilizer Measurement Circuits}}, 
  year={2025},
  volume={01},
  number={},
  pages={148-157},
  doi={10.1109/QCE65121.2025.00026}}

@misc{dem_trapping_sets,
    author       = {{Error-Correction-Lab}},
    title        = {{dem\_trapping\_sets}: Trapping-Set Enumeration for Circuit-Level Detector Error Models},
    year         = {2026},
    howpublished = {\url{https://github.com/Error-Correction-Lab/dem_trapping_sets}},
    note         = {GitHub repository}
}

@misc{chytas2026edge,
      title={{Edge-Based Anisotropic Decoding for Generalized Bicycle Codes}}, 
      author={Dimitris Chytas and Paul N. Fessatidis and Boulat A. Bash and Bane Vasić},
      year={2026},
      eprint={2605.03218},
      archivePrefix={arXiv},
}

\appendix 

\section{Structure of the DEM}
\label{sec:dem_structure}

In this Appendix, we describe the structure of the DEM in detail. For convenience, we only consider one type of detector, say $X$-type. Because of this we simply consider the $X$-stabilizer matrix $H_X$ and call it $H$, unless specified otherwise.

\paragraph{Symbolic form of the detector error model.}

A detector error model represents each elementary physical fault mechanism
by a binary column specifying the detectors flipped by that fault. For
\(r\) repeated syndrome-extraction rounds, let

\begin{equation*}
H_{\mathrm{DEM}}^{(r)}
=
\begin{bmatrix}
D^{(r)} & A^{(r)} & U^{(r)}
\end{bmatrix}
\end{equation*}

denote the detector matrix. Its columns are divided into three classes:
\(D^{(r)}\) contains faults originating on data qubits,
\(A^{(r)}\) contains faults that flip ancilla measurement outcomes without
propagating to the data, and \(U^{(r)}\) contains correlated faults produced
by error propagation through the syndrome-extraction circuit.

Let

\begin{equation*}
e
=
\begin{bmatrix}
e_D\\
e_A\\
e_U
\end{bmatrix}
\end{equation*}

be the binary vector indicating which elementary fault mechanisms occur.
The resulting detector-event vector is

\begin{equation*}
d
=
H_{\mathrm{DEM}}^{(r)}e
=
D^{(r)}e_D
+
A^{(r)}e_A
+
U^{(r)}e_U,
\end{equation*}

where all operations are over \(\mathbb{F}_2\).

Each column of \(H_{\mathrm{DEM}}^{(r)}\) corresponds to one
independent fault variable and is assigned a probability determined by
the underlying circuit-level noise model. Different physical mechanisms
may produce the same detector column, and are merged together if they produce the same logical action (if not, the DEM minimum distance is two).

To retain the logical action of each fault, one may similarly introduce
the observable matrix

\begin{equation*}
L_{\mathrm{DEM}}^{(r)}
=
\begin{bmatrix}
L_D^{(r)} & L_A^{(r)} & L_U^{(r)}
\end{bmatrix},
\end{equation*}

so that the logical-observable flips generated by \(e\) are

\begin{equation*}
\ell
=
L_{\mathrm{DEM}}^{(r)}e.
\end{equation*}

The complete symbolic detector error model is therefore specified by
the pair

\begin{equation*}
\left(
H_{\mathrm{DEM}}^{(r)},
L_{\mathrm{DEM}}^{(r)}
\right),
\end{equation*}

together with the probabilities assigned to its fault variables. In the
following, we describe separately the structure of the data, ancilla, and
hook-error blocks \(D^{(r)}\), \(A^{(r)}\), and \(U^{(r)}\).

\paragraph{Structure of \(D\).}

Let \(H\in\mathbb{F}_2^{m\times n}\) be the stabilizer matrix associated
with one family of checks, and let \(T\) be its Tanner graph. Each edge
of \(T\) represents a CNOT gate between a data qubit and the ancilla
used to measure the corresponding check. Since two CNOT gates sharing
a data or ancilla qubit cannot be executed simultaneously, the CNOT
schedule can be represented by a proper edge coloring

\begin{equation*}
    \sigma:E(T)\longrightarrow \{1,\ldots,t\},
\end{equation*}

where \(\sigma(e)\) specifies the time step at which the CNOT
corresponding to edge \(e\) is applied, and \(t\) is the number of CNOT
layers. For each variable node \(v\), denote the
ordered colors of its incident edges by

\begin{equation*}
c_{v,1}<c_{v,2}<\cdots<c_{v,d_v}.
\end{equation*}

The syndrome produced by a data fault changes only when such fault occurs after one or more CNOT gates involving that data qubit have been already processed. Hence, a variable node
of degree \(d_v\) generates \(d_v\) distinct variable nodes in the DEM.

Let
\[
    \mathcal{N}(v)
    =
    \left\{
        u\in\{1,\ldots,m\}:H_{u,v}=1
    \right\}
\]
be the set of checks acting on data qubit \(v\). For the \(i\)-th fault
class associated with \(v\), let \(c_{v,i}\) denote the time at which
the fault occurs. We partition \(\mathcal{N}(v)\) into
\begin{align}
    \mathcal{N}_{\mathrm{cur}}(v,i)
    &=
    \left\{
        u\in\mathcal{N}(v):
        \sigma(u,v)\geq c_{v,i}
    \right\},
    \\
    \mathcal{N}_{\mathrm{next}}(v,i)
    &=
    \left\{
        u\in\mathcal{N}(v):
        \sigma(u,v)<c_{v,i}
    \right\}.
\end{align}
Checks in \(\mathcal{N}_{\mathrm{cur}}(v,i)\) have not yet interacted
with \(v\) when the fault occurs and can therefore detect it in the
current round. Checks in
\(\mathcal{N}_{\mathrm{next}}(v,i)\) have already interacted with
\(v\), and detect the fault only in the following round.

Let \(s_{v,i},p_{v,i}\in\mathbb{F}_2^m\) be the indicator vectors of
\(\mathcal{N}_{\mathrm{cur}}(v,i)\) and
\(\mathcal{N}_{\mathrm{next}}(v,i)\), respectively. Since these two
sets form a partition of \(\mathcal{N}(v)\), it follows that
\begin{equation*}
    s_{v,i}+p_{v,i}=He_v,
\end{equation*}
where \(e_v\in\mathbb{F}_2^n\) is the unit vector associated with data
qubit \(v\).

Let

\begin{equation*}
N_D=\sum_{v=1}^{n}d_v
\end{equation*}

be the number of compressed data-fault classes in one round, and define

\begin{equation*}
S=
\begin{bmatrix}
s_{1,1} & \cdots & s_{v,i} & \cdots
\end{bmatrix},
\quad
P=
\begin{bmatrix}
p_{1,1} & \cdots & p_{v,i} & \cdots
\end{bmatrix}.
\end{equation*}

For \(r\) repeated syndrome-extraction rounds, the data part of the
detector matrix is

\begin{equation*}
D^{(r)}
=
\begin{bmatrix}
S      & 0      & \cdots & 0 \\
P      & S      & \ddots & \vdots \\
0      & P      & \ddots & 0 \\
\vdots & \ddots & \ddots & S \\
0      & \cdots & 0      & P
\end{bmatrix}.
\label{eq:data-dem-matrix}
\end{equation*}

Thus, every data-fault variable node has detector support on at
most two consecutive detector rounds. If the terminal detector layer
is not included, the final block row is removed.

A fault occurring after the final interaction of \(v\) in round \(q\)
has the same detector signature as a fault occurring before the first
interaction of \(v\) in round \(q+1\). This identification explains why
\(v\) generates \(d_v\), rather than \(d_v+1\), distinct DEM variables
per round.

\paragraph{Structure of \(A\).}

The matrix \(A\) describes faults that flip an ancilla measurement
outcome without propagating an error to the data. Since detectors compare
consecutive syndrome outcomes, an ancilla measurement fault in round
\(q\) flips the corresponding detectors in rounds \(q\) and \(q+1\).

For \(r\) repeated syndrome-extraction rounds, the ancilla part of the
detector matrix is therefore

\begin{equation*}
A^{(r)}
=
\begin{bmatrix}
I_m    & 0      & \cdots & 0 \\
I_m    & I_m    & \ddots & \vdots \\
0      & I_m    & \ddots & 0 \\
\vdots & \ddots & \ddots & I_m \\
0      & \cdots & 0      & I_m
\end{bmatrix}.
\label{eq:ancilla-dem-matrix}
\end{equation*}

\paragraph{Structure of \(U^{(r)}\).}

The matrix \(U^{(r)}\) describes correlated data errors generated by the
propagation of ancilla faults through the syndrome-extraction circuit.
These correlated errors are commonly referred to as \emph{hook errors}.

Hook errors detected by \(X\)-type detectors are \(Z\) errors originating
from the ancillas used to measure the \(Z\) checks. Their propagation
therefore depends on the CNOT ordering obtained from the coloring of
\(H_Z\), rather than \(H_X\).

Consider the ancilla associated with row \(u\) of \(H_Z\), and order the
data qubits in its support according to their CNOT times:
\begin{equation*}
    \sigma(u,v_{u,1})
    <
    \sigma(u,v_{u,2})
    <
    \cdots
    <
    \sigma(u,v_{u,d_u}).
\end{equation*}

During the measurement of a \(Z\) check, the data qubits act as the CNOT
controls and the ancilla acts as the target. The relevant propagation
rule is
\begin{equation*}
    Z_t \longmapsto Z_c Z_t.
\end{equation*}

Suppose that a \(Z\) fault occurs on the ancilla after its interaction
with \(v_{u,k-1}\) and before its interaction with \(v_{u,k}\), where
\(k=1\) denotes a fault occurring before the first CNOT. The fault is
then present during the remaining CNOTs and propagates a \(Z\) error to
the data qubits
\(v_{u,k},v_{u,k+1},\ldots,v_{u,d_u}\).

Let \(c_\ell\) select the columns of \(D^{(r)}\) associated with the
individual data errors generated by the \(\ell\)-th hook mechanism. The
corresponding detector column is
\begin{equation*}
    u_\ell^{\mathrm{prop}}
    =
    D^{(r)}c_\ell,
\end{equation*}
where the product is over \(\mathbb{F}_2\). Thus,
\(D^{(r)}c_\ell\) is the binary sum of the selected columns of
\(D^{(r)}\).

Collecting the vectors \(c_\ell\) into a matrix \(C^{(r)}\), the
propagated hook contribution is
\begin{equation*}
    U_{\mathrm{prop}}^{(r)}
    =
    D^{(r)}C^{(r)}.
    \label{eq:hook-data-factorization}
\end{equation*}

As for the elementary data-error matrix \(D^{(r)}\), the hook-error
matrix has a sliding-window structure. A hook fault occurring during one
syndrome-extraction round can change the detector outcomes associated
with that round and with the following round. Therefore,
\begin{equation*}
    U_{\mathrm{prop}}^{(r)}
    =
    \begin{bmatrix}
        S_{\mathrm{h}} & 0              & \cdots & 0 \\
        P_{\mathrm{h}} & S_{\mathrm{h}} & \ddots & \vdots \\
        0              & P_{\mathrm{h}} & \ddots & 0 \\
        \vdots         & \ddots         & \ddots & S_{\mathrm{h}} \\
        0              & \cdots         & 0      & P_{\mathrm{h}}
    \end{bmatrix},
    \label{eq:hook-dem-matrix}
\end{equation*}
where \(S_{\mathrm{h}}\) contains the detector changes in the round in
which the hook fault occurs, while \(P_{\mathrm{h}}\) contains the
detector changes that appear in the following round. The same blocks are
repeated at each round because the syndrome-extraction circuit is
repeated.

The analogous mechanism occurs for the \(X\) checks. In that case, the
ancilla acts as the CNOT control and
\begin{equation*}
    X_c \longmapsto X_c X_t.
\end{equation*}
An \(X\) fault on an \(X\)-check ancilla therefore propagates \(X\)
errors to the data qubits involved in the remaining CNOTs. These hook
errors are detected by \(Z\)-type detectors, and their propagation
depends on the coloring of \(H_X\).
Although a hook column can be written as a binary sum of elementary
data- and measurement-error columns, it must still be included as a
separate decoder variable: in this way, the decoder sees a ``whitened'' noise model, rather than a correlated one, even though alternative approaches are possible (see, for instance, Ref.~\cite{pacenti2025turbo}).

Adding the columns of \(U^{(r)}\) therefore presents the circuit-level
noise model to the belief-propagation decoder in terms of the underlying
physical fault mechanisms. Columns that are zero or duplicate existing
columns of \(D^{(r)}\) or \(A^{(r)}\) may be combined during DEM
compression.

\section{\texttt{dpl-search} algorithm}
\label{sec:dpl-search}

The \texttt{dpl-search} algorithm exhaustively enumerates LETS instances
within the range \(a\leq a_{\max}\) and
\(b\leq b_{\max}^{a}\). Starting from simple cycles, it recursively
constructs larger instances through dot, path, and lollipop expansions.
We denote by \(\mathcal{I}_{k}^{a,b}\) the set of instances in class
\((a,b)\) generated from an initial cycle containing \(k\) variable nodes.
Since the Tanner graph is bipartite, such a cycle has Tanner-graph length
\(2k\). If the Tanner graph has girth \(g\), the smallest initial cycle
therefore contains \(g/2\) variable nodes.
For each class \((a,b)\), the expansion table \(\mathrm{EX}(a,b)\) specifies
whether a dot expansion must be applied, the path parameters \(m\) to be
considered, and the lollipop pairs \((m,c)\). The table is generated using
Algorithm~1 of Ref.~\cite{bani_irregular}, given the bounds
\(b_{\max}^{a}\) and the variable-node degrees appearing in the Tanner graph.
The search includes classes with \(b=0\). These instances are retained but
are terminal under the standard DPL expansions, since dot and path expansions
require at least two checks in \(\mathcal{N}_o(S)\), while a lollipop
expansion requires at least one.
Because the Tanner graph of a circuit-level detector error model is generally
irregular, every generated support is explicitly checked before being
retained. In particular, all checks in the induced subgraph must have degree
one or two, and every variable node must be adjacent to at least two checks
in \(\mathcal{N}_e(S)\).

\begin{revtexalgorithm}[h!]
\algcaption{\texttt{dpl-search}}
\label{alg:lets-exhaustive-search}
\begin{algorithmic}[1]
\REQUIRE Tanner graph $G$ with girth $g$; maximum size $a_{\max}$;
bounds $\{b_{\max}^{g/2},\ldots,b_{\max}^{a_{\max}}\}$;
expansion table $\mathrm{EX}$
\ENSURE Set $\mathcal{I}$ containing all LETS instances satisfying
$a\leq a_{\max}$ and $b\leq b_{\max}^{a}$

\STATE $\mathcal{I}\gets\emptyset$

\FOR{$k=g/2,\ldots,a_{\max}$}

    \FOR{$a=k,\ldots,a_{\max}$}
        \FOR{$b=0,\ldots,b_{\max}^{a}$}
            \STATE $\mathcal{I}_{k}^{a,b}\gets\emptyset$
        \ENDFOR
    \ENDFOR

    \STATE $\{\mathcal{C}_{k}^{b}\}_{b=0}^{b_{\max}^{k}}
    \gets
    \texttt{enumerate\_cycles}
    (G,k,b_{\max}^{k})$

    \STATE $\mathcal{C}_{k}
    \gets
    \displaystyle\bigcup_{b=0}^{b_{\max}^{k}}
    \mathcal{C}_{k}^{b}$

    \FOR{$b=0,\ldots,b_{\max}^{k}$}
        \STATE $\mathcal{I}_{k}^{k,b}
        \gets
        \mathcal{I}_{k}^{k,b}
        \cup
        \mathcal{C}_{k}^{b}$

        \STATE $\mathcal{I}
        \gets
        \mathcal{I}
        \cup
        \mathcal{C}_{k}^{b}$
    \ENDFOR

\ENDFOR

\FOR{$k=g/2,\ldots,a_{\max}$}

    \FOR{$a=k,\ldots,a_{\max}-1$}

        \FOR{$b=0,\ldots,b_{\max}^{a}$}

            \IF{$\mathcal{I}_{k}^{a,b}\neq\emptyset$}

                \IF{$\texttt{dot}\in\mathrm{EX}(a,b)$}

                    \STATE $\{\mathcal{D}^{s}\}_{s=0}^{b_{\max}^{a+1}}
                    \gets
                    \texttt{dot\_expansion}
                    \left(
                        G,
                        \mathcal{I}_{k}^{a,b},
                        b_{\max}^{a+1}
                    \right)$

                    \FOR{$s=0,\ldots,b_{\max}^{a+1}$}

                        \STATE $\mathcal{I}_{k}^{a+1,s}
                        \gets
                        \mathcal{I}_{k}^{a+1,s}
                        \cup
                        \mathcal{D}^{s}$

                        \STATE $\mathcal{I}
                        \gets
                        \mathcal{I}
                        \cup
                        \mathcal{D}^{s}$

                    \ENDFOR

                \ENDIF

                \FOR{each $\texttt{pa}_{m}\in\mathrm{EX}(a,b)$}

                    \IF{$a+m\leq a_{\max}$}

                        \STATE $\{\mathcal{P}_{m}^{s}\}
                        _{s=0}^{b_{\max}^{a+m}}
                        \gets
                        \texttt{path\_expansion}
                        \left(
                            G,
                            \mathcal{I}_{k}^{a,b},
                            m,
                            b_{\max}^{a+m}
                        \right)$

                        \FOR{$s=0,\ldots,b_{\max}^{a+m}$}

                            \STATE $\mathcal{I}_{k}^{a+m,s}
                            \gets
                            \mathcal{I}_{k}^{a+m,s}
                            \cup
                            \mathcal{P}_{m}^{s}$

                            \STATE $\mathcal{I}
                            \gets
                            \mathcal{I}
                            \cup
                            \mathcal{P}_{m}^{s}$

                        \ENDFOR

                    \ENDIF

                \ENDFOR

                \FOR{each $\texttt{lo}_{m}^{c}\in\mathrm{EX}(a,b)$}

                    \IF{$a+m\leq a_{\max}$}

                        \STATE $\{\mathcal{L}_{m,c}^{s}\}
                        _{s=0}^{b_{\max}^{a+m}}
                        \gets
                        \texttt{lollipop\_expansion}
                        \left(
                            G,
                            \mathcal{I}_{k}^{a,b},
                            \mathcal{C}_{c},
                            m,
                            c,
                            b_{\max}^{a+m}
                        \right)$

                        \FOR{$s=0,\ldots,b_{\max}^{a+m}$}

                            \STATE $\mathcal{I}_{k}^{a+m,s}
                            \gets
                            \mathcal{I}_{k}^{a+m,s}
                            \cup
                            \mathcal{L}_{m,c}^{s}$

                            \STATE $\mathcal{I}
                            \gets
                            \mathcal{I}
                            \cup
                            \mathcal{L}_{m,c}^{s}$

                        \ENDFOR

                    \ENDIF

                \ENDFOR

            \ENDIF

        \ENDFOR

    \ENDFOR

\ENDFOR

\STATE \textbf{return} $\mathcal{I}$
\end{algorithmic}
\end{revtexalgorithm}

\subsection{Enumeration of simple cycles}

Algorithm~\ref{alg:cyc-srch}, denoted by
\texttt{enumerate\_cycles}, enumerates all chordless simple cycles containing
\(k\) variable nodes, or equivalently having Tanner-graph length \(2k\).

For each cycle \(Q\), let \(S=\operatorname{Var}(Q)\) be its variable-node
support and \(\operatorname{Chk}(Q)\) its check-node set. The candidate is
retained only if \(|S|=k\),
\(\mathcal{N}_e(S)=\operatorname{Chk}(Q)\), and
\(|\mathcal{N}_o(S)|\leq b_{\max}^{k}\). The equality
\(\mathcal{N}_e(S)=\operatorname{Chk}(Q)\) excludes chords and additional
degree-two connections between nonconsecutive cycle variables.

The complete induced subgraph is then checked explicitly. Valid supports are
inserted into \(\mathcal{I}_{k}^{k,|\mathcal{N}_o(S)|}\). Since the same
cycle may be found from different starting nodes or traversal directions, the
collections are stored as sets and duplicate supports are removed
automatically.

\begin{revtexalgorithm}
\algcaption{\texttt{enumerate\_cycles}}
\label{alg:cyc-srch}
\begin{algorithmic}[1]
\REQUIRE Tanner graph $G=(V\cup C,E)$; cycle size $k$;
bound $b_{\max}^{k}$
\ENSURE Sets
$\{\mathcal{I}_{k}^{k,0},\ldots,
\mathcal{I}_{k}^{k,b_{\max}^{k}}\}$
containing all chordless simple-cycle LETS instances with $k$ variable nodes

\FOR{$b=0,\ldots,b_{\max}^{k}$}
    \STATE $\mathcal{I}_{k}^{k,b}\gets\emptyset$
\ENDFOR

\FOR{each simple cycle $Q$ in $G$ with Tanner-graph length $2k$}

    \STATE $S\gets\operatorname{Var}(Q)$
    \STATE $b\gets|\Gamma_o(S)|$

    \STATE $\texttt{elementary}\gets
    \left[
        \deg_{G(S)}(c)\in\{1,2\},
        \ \forall c\in\Gamma(S)
    \right]$

    \STATE $\texttt{leafless}\gets
    \left[
        |\Gamma(v)\cap\Gamma_e(S)|\geq 2,
        \ \forall v\in S
    \right]$

    \STATE $\texttt{chordless}\gets
    \left[
        \Gamma_e(S)=\operatorname{Chk}(Q)
    \right]$

    \IF{$b\leq b_{\max}^{k}$
        and $\texttt{elementary}$
        and $\texttt{leafless}$
        and $\texttt{chordless}$}

        \STATE $\mathcal{I}_{k}^{k,b}
        \gets
        \mathcal{I}_{k}^{k,b}\cup\{S\}$

    \ENDIF

\ENDFOR

\STATE \textbf{return}
$\{\mathcal{I}_{k}^{k,0},\ldots,
\mathcal{I}_{k}^{k,b_{\max}^{k}}\}$
\end{algorithmic}
\end{revtexalgorithm}

\subsection{Dot expansion}

Algorithm~\ref{alg:dot-srch}, denoted by
\texttt{dot\_expansion}, adds one variable node to a support
\(S\in\mathcal{I}_{k}^{a,b}\).
A node \(v\notin S\) is considered only if
\(|\mathcal{N}(v)\cap\mathcal{N}_o(S)|\geq 2\) and
\(\mathcal{N}(v)\cap\mathcal{N}_e(S)=\emptyset\). The first condition
ensures that \(v\) is connected to the parent through at least two
unsatisfied checks, while the second prevents an existing satisfied check
from acquiring induced degree three.
The expanded support is \(S'=S\cup\{v\}\), with \(|S'|=a+1\). If
\(t=|\mathcal{N}(v)\cap\mathcal{N}_o(S)|\) and
\(q=|\mathcal{N}(v)\setminus\mathcal{N}(S)|\), then
\(|\mathcal{N}_o(S')|=b-t+q\). Equivalently, since
\(\deg_G(v)=t+q\), we have
\(|\mathcal{N}_o(S')|=b+\deg_G(v)-2t\).
The support is retained only if
\(|\mathcal{N}_o(S')|\leq b_{\max}^{a+1}\) and the complete induced
subgraph satisfies the required structural conditions.

\begin{revtexalgorithm}
\algcaption{\texttt{dot\_expansion}}
\label{alg:dot-srch}
\begin{algorithmic}[1]
\REQUIRE Tanner graph $G=(V\cup C,E)$; LETS instances
$\mathcal{I}_{k}^{a,b}$; bound $b_{\max}^{a+1}$
\ENSURE Sets
$\{\mathcal{D}^{0},\ldots,\mathcal{D}^{b_{\max}^{a+1}}\}$
containing the LETS instances generated by dot expansion

\FOR{$s=0,\ldots,b_{\max}^{a+1}$}
    \STATE $\mathcal{D}^{s}\gets\emptyset$
\ENDFOR

\FOR{each LETS instance $S\in\mathcal{I}_{k}^{a,b}$}

    \STATE $\mathcal{V}_{\mathrm{cand}}\gets
    \left\{
        v\in V\setminus S:
        |\Gamma(v)\cap\Gamma_o(S)|\geq 2
        \text{ and }
        \Gamma(v)\cap\Gamma_e(S)=\emptyset
    \right\}$

    \FOR{each variable node $v\in\mathcal{V}_{\mathrm{cand}}$}

        \STATE $S'\gets S\cup\{v\}$
        \STATE $b'\gets|\Gamma_o(S')|$

        \STATE $\texttt{elementary}\gets
        \left[
            \deg_{G(S')}(c)\in\{1,2\},
            \ \forall c\in\Gamma(S')
        \right]$

        \STATE $\texttt{leafless}\gets
        \left[
            |\Gamma(u)\cap\Gamma_e(S')|\geq 2,
            \ \forall u\in S'
        \right]$

        \IF{$b'\leq b_{\max}^{a+1}$
            and $\texttt{elementary}$
            and $\texttt{leafless}$}

            \STATE $\mathcal{D}^{b'}
            \gets
            \mathcal{D}^{b'}\cup\{S'\}$

        \ENDIF

    \ENDFOR

\ENDFOR

\STATE \textbf{return}
$\{\mathcal{D}^{0},\ldots,\mathcal{D}^{b_{\max}^{a+1}}\}$
\end{algorithmic}
\end{revtexalgorithm}

\subsection{Path expansion}

Algorithm~\ref{alg:path-srch}, denoted by
\texttt{path\_expansion}, connects two distinct checks in
\(\mathcal{N}_o(S)\) through a new alternating path.
The parameter \(m\) denotes the number of new variable nodes. A path expansion therefore has the form
\(p=(c_0,v_1,c_1,\ldots,c_{m-1},v_m,c_m)\), where
\(c_0,c_m\in\mathcal{N}_o(S)\) are distinct. The path has Tanner-graph
length \(2m\) and introduces the variable set
\(X_p=\operatorname{Var}(p)=\{v_1,\ldots,v_m\}\).
The path is admissible only if \(X_p\cap S=\emptyset\) and
\({\mathcal{N}(X_p)\cap\mathcal{N}(S)=\{c_0,c_m\}}\). Thus, it intersects
the parent only through the two selected endpoint checks. The internal checks
\(c_1,\ldots,c_{m-1}\) must be distinct, lie outside
\(\mathcal{N}(S)\), and be the only checks adjacent to two variables in
\(X_p\). These conditions exclude repeated nodes, chords, branches, and
additional connections to the parent.
The expanded support is \(S'=S\cup X_p\), with \(|S'|=a+m\). It is
retained only if \(|\mathcal{N}_o(S')|\leq b_{\max}^{a+m}\) and the
complete induced subgraph satisfies the required structural conditions.

\begin{revtexalgorithm}
\algcaption{\texttt{path\_expansion}}
\label{alg:path-srch}
\begin{algorithmic}[1]
\REQUIRE Tanner graph $G=(V\cup C,E)$; LETS instances
$\mathcal{I}_{k}^{a,b}$; path parameter $m$;
bound $b_{\max}^{a+m}$
\ENSURE Sets
$\{\mathcal{P}_{m}^{0},\ldots,
\mathcal{P}_{m}^{b_{\max}^{a+m}}\}$
containing the LETS instances generated by path expansion

\FOR{$s=0,\ldots,b_{\max}^{a+m}$}
    \STATE $\mathcal{P}_{m}^{s}\gets\emptyset$
\ENDFOR

\FOR{each LETS instance $S\in\mathcal{I}_{k}^{a,b}$}

    \FOR{each unordered pair of distinct check nodes
    $c_i,c_j\in\Gamma_o(S)$}

        \FOR{each simple alternating path
        $p=(c_i,v_1,r_1,v_2,\ldots,r_{m-1},v_m,c_j)$ in $G$}

            \STATE $X_p\gets\operatorname{Var}(p)
            =\{v_1,\ldots,v_m\}$

            \STATE $\mathcal{R}_p\gets
            \{r_1,\ldots,r_{m-1}\}$

            \IF{$X_p\cap S=\emptyset$
            and
            $\Gamma(X_p)\cap\Gamma(S)=\{c_i,c_j\}$
            and
            $\Gamma_e(X_p)=\mathcal{R}_p$}

                \STATE $S'\gets S\cup X_p$
                \STATE $b'\gets|\Gamma_o(S')|$

                \STATE $\texttt{elementary}\gets
                \left[
                    \deg_{G(S')}(c)\in\{1,2\},
                    \ \forall c\in\Gamma(S')
                \right]$

                \STATE $\texttt{leafless}\gets
                \left[
                    |\Gamma(v)\cap\Gamma_e(S')|\geq 2,
                    \ \forall v\in S'
                \right]$

                \IF{$|S'|=a+m$
                and
                $b'\leq b_{\max}^{a+m}$
                and
                $\texttt{elementary}$
                and
                $\texttt{leafless}$}

                    \STATE $\mathcal{P}_{m}^{b'}
                    \gets
                    \mathcal{P}_{m}^{b'}\cup\{S'\}$

                \ENDIF

            \ENDIF

        \ENDFOR

    \ENDFOR

\ENDFOR

\STATE \textbf{return}
$\{\mathcal{P}_{m}^{0},\ldots,
\mathcal{P}_{m}^{b_{\max}^{a+m}}\}$
\end{algorithmic}

\end{revtexalgorithm}

\subsection{Lollipop expansion}

Algorithm~\ref{alg:lolli-srch}, denoted by
\texttt{lollipop\_expansion}, connects a support
\(S\in\mathcal{I}_{k}^{a,b}\) to a simple cycle \(C\) containing \(c\)
variable nodes through a path, or \textit{stem}.
The parameter \(m\) denotes the total number of variable nodes introduced by
the expansion. Defining \(d=m+1-c\), the cycle contributes \(c\) variables
and the stem contributes \(d-1\), so that \(c+d-1=m\).
When \(d=1\), there is no stem variable and \(m=c\). The supports must satisfy
\(S\cap C=\emptyset\), and the parent and cycle must share exactly one check
\(q\in\mathcal{N}_o(S)\cap\mathcal{N}_o(C)\), namely
\(\mathcal{N}(S)\cap\mathcal{N}(C)=\{q\}\). The expanded support is then
\(S'=S\cup C\).
When \(d>1\), the stem has the form
\(p=(q_0,x_1,r_1,\ldots,r_{d-2},x_{d-1},q_1)\), where
\(q_0\in\mathcal{N}_o(S)\) and \(q_1\in\mathcal{N}_o(C)\). Its variable
set is \(X_p=\{x_1,\ldots,x_{d-1}\}\), and its Tanner-graph length is
\(2(d-1)\).
The parent and cycle must satisfy \(S\cap C=\emptyset\) and
\(\mathcal{N}(S)\cap\mathcal{N}(C)=\emptyset\). Moreover,
\(X_p\cap(S\cup C)=\emptyset\),
\(\mathcal{N}(X_p)\cap\mathcal{N}(S)=\{q_0\}\), and
\(\mathcal{N}(X_p)\cap\mathcal{N}(C)=\{q_1\}\). These conditions prevent
the stem from intersecting the parent or cycle through an additional variable
node, satisfied check, or unsatisfied check.
The internal stem checks \(r_1,\ldots,r_{d-2}\) must be distinct and must be
the only checks adjacent to two variables in \(X_p\). The expanded support is
\(S'=S\cup C\cup X_p\), with \(|S'|=a+m\).
In both cases, the candidate is retained only if
\(|\mathcal{N}_o(S')|\leq b_{\max}^{a+m}\) and the complete induced
subgraph satisfies the required structural conditions.

\begin{revtexalgorithm}[h!]
\algcaption{\texttt{lollipop\_expansion}}
\label{alg:lolli-srch}
\begin{algorithmic}[1]
\REQUIRE Tanner graph $G=(V\cup C,E)$; LETS instances
$\mathcal{I}_{k}^{a,b}$; simple-cycle instances $\mathcal{C}_{c}$;
lollipop parameters $m$ and $c$; bound $b_{\max}^{a+m}$
\ENSURE Sets
$\{\mathcal{L}_{m,c}^{0},\ldots,
\mathcal{L}_{m,c}^{b_{\max}^{a+m}}\}$
containing the LETS instances generated by lollipop expansion

\STATE $d\gets m+1-c$

\FOR{$s=0,\ldots,b_{\max}^{a+m}$}
    \STATE $\mathcal{L}_{m,c}^{s}\gets\emptyset$
\ENDFOR

\FOR{each LETS instance $S\in\mathcal{I}_{k}^{a,b}$}

    \FOR{each simple-cycle instance $C\in\mathcal{C}_{c}$ such that
    $S\cap C=\emptyset$}

        \IF{$d=1$}

            \IF{$\Gamma(S)\cap\Gamma(C)=\{q\}$ for some
            $q\in\Gamma_o(S)\cap\Gamma_o(C)$}

                \STATE $S'\gets S\cup C$
                \STATE $b'\gets|\Gamma_o(S')|$

                \STATE $\texttt{elementary}\gets
                \left[
                    \deg_{G(S')}(r)\in\{1,2\},
                    \ \forall r\in\Gamma(S')
                \right]$

                \STATE $\texttt{leafless}\gets
                \left[
                    |\Gamma(v)\cap\Gamma_e(S')|\geq 2,
                    \ \forall v\in S'
                \right]$

                \IF{$|S'|=a+m$
                and $b'\leq b_{\max}^{a+m}$
                and $\texttt{elementary}$
                and $\texttt{leafless}$}

                    \STATE $\mathcal{L}_{m,c}^{b'}
                    \gets
                    \mathcal{L}_{m,c}^{b'}\cup\{S'\}$

                \ENDIF

            \ENDIF

        \ELSE

            \IF{$\Gamma(S)\cap\Gamma(C)=\emptyset$}

                \FOR{each simple alternating path
                \[
                    p=
                    (q_0,x_1,r_1,x_2,\ldots,
                    r_{d-2},x_{d-1},q_1)
                \]
                in $G$, where
                $q_0\in\Gamma_o(S)$ and
                $q_1\in\Gamma_o(C)$}

                    \STATE $X_p\gets\operatorname{Var}(p)
                    =\{x_1,\ldots,x_{d-1}\}$

                    \STATE $\mathcal{R}_p
                    \gets\{r_1,\ldots,r_{d-2}\}$

                    \IF{$X_p\cap(S\cup C)=\emptyset$
                    and
                    $\Gamma(X_p)\cap\Gamma(S)=\{q_0\}$
                    and
                    $\Gamma(X_p)\cap\Gamma(C)=\{q_1\}$
                    and
                    $\Gamma_e(X_p)=\mathcal{R}_p$}

                        \STATE $S'\gets S\cup C\cup X_p$
                        \STATE $b'\gets|\Gamma_o(S')|$

                        \STATE $\texttt{elementary}\gets
                        \left[
                            \deg_{G(S')}(r)\in\{1,2\},
                            \ \forall r\in\Gamma(S')
                        \right]$

                        \STATE $\texttt{leafless}\gets
                        \left[
                            |\Gamma(v)\cap\Gamma_e(S')|\geq 2,
                            \ \forall v\in S'
                        \right]$

                        \IF{$|S'|=a+m$
                        and $b'\leq b_{\max}^{a+m}$
                        and $\texttt{elementary}$
                        and $\texttt{leafless}$}

                            \STATE $\mathcal{L}_{m,c}^{b'}
                            \gets
                            \mathcal{L}_{m,c}^{b'}
                            \cup\{S'\}$

                        \ENDIF

                    \ENDIF

                \ENDFOR

            \ENDIF

        \ENDIF

    \ENDFOR

\ENDFOR

\STATE \textbf{return}
$\{\mathcal{L}_{m,c}^{0},\ldots,
\mathcal{L}_{m,c}^{b_{\max}^{a+m}}\}$
\end{algorithmic}
\end{revtexalgorithm}

\subsection{Complete exhaustive-search procedure}

Algorithm~\ref{alg:lets-exhaustive-search} first invokes
\texttt{enumerate\_cycles} for
\(k=g/2,g/2+1,\ldots,a_{\max}\). The resulting cycles are classified by
their values of \(k\) and \(b\) and inserted into the corresponding
collections \(\mathcal{I}_{k}^{k,b}\).
For each initial cycle size \(k\), the structures are then processed in
increasing order of \(a\). For every class
\(0\leq b\leq b_{\max}^{a}\), the entry \(\mathrm{EX}(a,b)\) specifies
which expansions must be applied. The classes with \(b=0\) are included,
although their expansion-table entries are empty.
Each call to \texttt{dot\_expansion}, \texttt{path\_expansion}, or
\texttt{lollipop\_expansion} returns collections indexed by the resulting
value of \(b\). These collections are immediately accumulated into the
corresponding \(\mathcal{I}_{k}^{a',b'}\) using set union. In particular,
the output of one parent, path parameter, or lollipop pair must not overwrite
the output of another.
Every candidate \(S'\) is classified by
\((|S'|,|\mathcal{N}_o(S')|)\) and discarded if
\(|S'|>a_{\max}\) or
\(|\mathcal{N}_o(S')|>b_{\max}^{|S'|}\). It is retained only after the
complete induced subgraph has been validated.
Since every expansion strictly increases the support size, the search
proceeds from smaller to larger values of \(a\) and terminates once
\(a_{\max}\) is reached.

\end{document}